\documentclass[preprint,aps,amsmath,nofootinbib]{revtex4}
\usepackage{graphics,color,array,dcolumn}
\usepackage{graphicx}
\usepackage{calc}

\usepackage{amsmath}
\usepackage{amssymb}
\usepackage{xspace}
\allowdisplaybreaks[1]
\newlength{\figurewidth}
\newcommand{\beq}{\begin{equation}}
\newcommand{\eeq}{\end{equation}}
\newcommand{\bea}{\begin{eqnarray}}
\newcommand{\eea}{\end{eqnarray}}
\newcommand{\ba}{\begin{array}}
\newcommand{\ea}{\end{array}}
\newcommand{\bg}{\bar{g}}
\newcommand{\mn}{{\mu\nu}}
\newcommand{\pt}{\partial}
\newcommand{\pd}{(2\pi)^d}
\newcommand{\al}{\alpha}
\newcommand{\bt}{\beta}
\newcommand{\g}{\gamma}
\newcommand{\ep}{\epsilon}
\newcommand{\ta}{\theta}
\newcommand{\lam}{\lambda}

\newcommand{\G}{\Gamma}

\newcommand{\de}{\delta}
\newcommand{\D}{\Delta}
\newcommand{\sg}{\sigma}
\newcommand{\kp}{\kappa}
\newcommand{\bnb}{\bar{\nabla}}
\makeatother
\begin{document}
%
\title{Running Couplings in Quantum Theory of Gravity Coupled with Gauge Fields}
\setlength{\figurewidth}{\columnwidth}
%
\author{Gaurav Narain}
\email{gaunarain@imsc.res.in}
\author{Ramesh Anishetty}
\email{ramesha@imsc.res.in}
\affiliation{
The Institute of Mathematical Sciences, 
CIT Campus, Taramani, Chennai 600113, India.}
%
%
\begin{abstract}
In this paper we study the coupled system of non-abelian 
gauge fields with higher-derivative gravity. Charge renormalization 
is investigated in this coupled system. It is found that the leading 
term in the gauge coupling beta function comes due to interaction of gauge 
fields with gravitons. This is shown to be a universal quantity in the sense that
it doesn't depend on the gauge coupling and the gauge group, 
but may depend on the other couplings of the action (gravitational and matter). 
The coupled system is studied at one-loop. It is found that the leading term 
of gauge beta function is zero at one-loop in four dimensions. The effect of gauge fields
on the running of gravitational couplings is investigated. 
The coupled system of gauge field with higher-derivative 
gravity is shown to satisfy unitarity when quantum corrections
are taken in to account. Moreover, it is found that
Newton constant goes to zero at short distances.
In this renormalizable and unitary theory of gauge field 
coupled with higher-derivative gravity, the leading term of
the gauge beta function, found to be universal for all
gauge groups, is further studied in more detail by 
isolating it in the context of abelian gauge theories coupled 
with gravity in four dimensions. Using self-duality of abelian gauge theories in four 
dimensions, this term of the gauge beta function is shown to be zero to all
loops. This is found to be independent of the gravity action,
regularization scheme and gauge fixing condition. An explicit one-loop computation
for arbitrary gravity action further demonstrates the vanishing of this term 
in the gauge beta function in four dimensions, independent of the regularization 
scheme and gauge fixing condition. Consequences of this are discussed.
\end{abstract}

\maketitle
%
%

\section{Introduction}
\label{intro}

Matter that has been observed so far can be very well described 
theoretically using the standard model of particle physics, which has been very accurately 
tested in accelerator experiments. With the discovery of a scalar particle at 125 GeV 
(expected so far to be Higgs particle), the standard model of particle physics is seen 
as an ultimate theory of nature which is capable of describing it to very high energies,
may be all the way up to Planck scale \cite{EliasMiro,Degrassi,Masina}. 
It theoretically describes three of the four known forces 
of nature: Electromagnetic, Weak and Strong forces. These forces are described using the 
gauge fields whose quanta acts as force carriers between matter. 
The gauge fields describing the electromagnetic force is abelian in nature, 
while the gauge fields describing the weak and strong forces are 
non-abelian in nature. All gauge fields couple with matter but in case
of non-abelian gauge theories, the fields also have self coupling. The 
strength of these couplings are described using parameters, which in 
a quantum theory changes with energy scale (a generic feature of 
any quantum field theory), meaning that at different energy scales the 
strength of interaction is different, unlike in classical theories where strength 
of interactions remain fixed. Witnessing such theoretically predicted running 
of coupling parameters experimentally further justifies the methods 
and tools used to study quantum field theories. This running of gauge couplings 
can be extrapolated to very high energies as standard model of 
particle physics remains well defined all the way up to Planck scale. 
It is found that for non-abelian gauge 
couplings the running is such that the coupling parameter tends to zero at high 
enough energies (called asymptotic freedom), while for abelian gauge coupling
the running leads to a singularity at a particular energy at which the coupling blows 
up (Landau singularity). 

Theoretically these phenomenas can be studied in perturbation 
theory using Feynman path-integrals, which are sum over 
the phase-space configuration - each configuration being weighed 
by a phase. Perturbative quantum field theories 
can be studied very accurately using the path-integral techniques 
as long as the running coupling parameter remains small, in which 
case the observables obtained can be reliably estimated. In Standard Model (SM),
perturbation theory is well defined all the way up to Planck scale,
as the Landau singularity of the abelian gauge coupling (which is a 
source of worry) occurs way beyond the Planck scale 
\cite{EliasMiro,Degrassi,Masina}. On the other hand
non-abelian gauge coupling witnesses asymptotic freedom in the 
high energy regime thereby guaranteeing the validity of perturbation 
theory. The problem with abelian gauge coupling is evaded in
Grand Unified Theories (GUT) by considering a bigger symmetry group of the matter fields 
\cite{GUT, GUTunity, SUSYunity}. In these scenarios there is enhanced 
symmetry at high energy scales greater than $10^{16}$ GeV, which 
breaks down at $10^{16}$ GeV to the known standard model symmetry group.
This scenario overcomes the problem of Landau singularity in a very 
elegant way by unifying all the standard model gauge couplings 
at $10^{16}$ GeV \cite{GUT, GUTunity, SUSYunity}. 
However in the absence of such enhanced symmetry groups (as there is 
no experimental signature so far confirming this possibility), 
problem of Landau singularity for abelian gauge coupling persist and it 
appears as a blot on the beautiful theory of standard model.
Moreover in the absence of GUT scenarios, standard model 
ultimately enters a non-perturbative domain, as the abelian 
gauge coupling becomes large near the Landau pole 
signaling the breakdown of the perturbation theory. 
GUT scenarios may seem elegant and beautiful solution to 
overcome the problem, but it not a necessity. In its absence
the validity of the SM is limited and the perturbative path-integral
is well defined only up to finite albeit larger than Plank energy scale. 

Gravitational attractive force between two particles (charged or
uncharged) becomes comparable to the electromagnetic, weak 
or strong force at the Planck scale. So far in studying SM,
we have ignored gravitational interactions all the way up to 
Planck scale. This is justified as gravitational force is quite 
weak compared to other three forces at scales below Planck 
energy. But certainly it should not be ignored near or at Planck scale, 
where gravitational effects are significantly important and 
would contribute equally to any of the quantum effects taking place at 
that energy scale. Also, the standard model experiments that 
are performed today are local events, where background 
curvature is irrelevant, as its effects are negligible. However parts 
of background spacetime, where curvature blows up, curvature 
effects become large and cannot be ignored. These occur near the 
singularities like the one in black holes. 
Places like these with large curvature, leads to particle creation.
This is analogous to the Schwinger mechanism \cite{Schwinger1} 
witnessed when strong electric field results in particle 
pair creation. A correct theoretical description of this is achieved when the 
full SM coupled with Einstein-Hilbert gravity is studied. 
Studying quantum matter fields on curved background provides 
an insight in to how the the standard model phenomenas will get 
corrected when background curvature is taken in to account?
Due to back reaction such effects will also modify the background 
gravitational field and leads to higher-derivative type gravitational 
interactions, signaling the non-renormalizabilty of the SM on 
a curved background. Still, such studies are important in their 
own light. In a complete quantum picture it is required that all fields present in 
the theory be quantized. This when attempted over the coupled system 
of SM with Einstein-Hilbert gravity leads to plague of ultraviolet 
divergences, where at each order of perturbation theory new 
counter-terms appear which cannot be absorbed in the previous ones, 
indicating one to conclude that the coupled system is non-renormalizable.
However, it is not a completely devastating situation, as one can still study in this 
coupled system how the low energy physics gets modified due to 
quantum nature of the gravitational field and whether such corrections
will have any observable consequences? This is the attitude taken 
towards these studies which goes under the name of Effective Field Theory
\cite{Donoghue1, Donoghue2, Donoghue3}.
However such effects cannot be trusted at very high energies (like Planck
scale), where a knowledge of sensible quantum gravity is required.

Running of standard model couplings is one place where effects are 
quantum gravity are likely to affect their behavior, because the presence
of new degree of freedom in the system modifies the behavior of running 
couplings, as the virtual particles of these new species are generated whose 
contributions are not negligible near Planck energy scale. In case of gauge theories,
such contributions are expected to arise and may modify the high energy 
behavior of running gauge couplings namely asymptotic freedom and 
Landau singularity. This effect if present will be hardly noticeable 
at low energies due to Planck mass suppression, but would certainly 
affect the delicate unification of the gauge couplings at GUT scale in the 
supersymmetric GUT theories \cite{GUT, GUTunity, SUSYunity}.
Such unifications being very delicate and sensitive to input 
parameters, are very easily affected due to any minor modification
to the gauge beta functions coming due to quantum gravity effects.
If the presence of virtual graviton were 
to spoil the unification in such a manner, then it is very disturbing 
as the unification of Standard model gauge couplings is a 
necessary prediction coming out from any realistic GUT theories. 
In the absence of GUT (which is not a necessity to solve the 
problems of SM), the running of abelian gauge coupling can't be 
extrapolated to very high energies as it runs in to problem (Landau 
singularity). On the other hand in the case of non-abelian gauge 
couplings the running can be extrapolated to arbitrarily high energies. 
It is difficult to conceive that such runnings remains unaffected 
even at very high energies for example Planck scale, where quantum 
gravity effects are important. It therefore becomes crucial to examine 
and study how such runnings gets altered under the influence of 
quantum gravity effects?

Effects of quantum gravity on the running of standard model couplings have 
been studied in the past. This has been mostly done in the context of standard model 
of particle physics coupled with Einstein-Hilbert (EH) gravity 
\cite{thooft1, Deser1, Deser2, Deser3, Deser4, Deser5, Deser6}. This is a 
non-renormalizable system, thus any computation that has been done in this 
context is studied within the framework of Effective field 
theory \cite{Donoghue1, Donoghue2, Donoghue3}. 
Effects of quantum gravity on the renormalization of charge were 
first discussed in \cite{Deser1, Deser4, Deser5, Deser6} within perturbation theory using 
$4-\ep$ dimensional regularization scheme \cite{thooft2}, and concluded that at 
one-loop there are no quantum gravity correction to the beta function 
of the gauge couplings. This result was often suspected to be a 
consequence of the massless nature of graviton and gluons/photons,
and the way dimensional regularization handles the quadratic divergences.
This problem was re-examined by using a momentum cutoff with a
$R_\xi$ type of gauge fixing condition at one-loop in \cite{Robinson}.
They came to the conclusion that the beta function of the gauge coupling 
gets a nonzero quantum gravity correction signaling asymptotic freedom 
of all gauge couplings even those which hits Landau singularity at high
energies. This also showed that the coupling vanishes as power law as
opposed to logarithmic well before Planck energies. The authors used 
momentum cutoff as it is able to see quadratic divergences present in the 
theory, which dimensional regularization misses in four dimensions.

However this result came under criticism and was re-studied by several 
authors using various ways in different gauge choices or regularization 
scheme and refuted the result of \cite{Robinson}. The investigations which 
find that there is no quantum gravity correction to charge renormalization 
are the following: using momentum cutoff and a harmonic type 
gauge fixing condition, quadratic divergences were studied in \cite{Pietrykowski}; using 
dimensional regularization with a gauge independent formulation 
of effective action in \cite{Vilkovisky1984,Toms:2007}; in \cite{Rodigast2008} the 
problem was studied using feynman diagram technique within both 
momentum and dimensional regularization scheme; using 
functional renormalization group by choosing a symmetry preserving 
gauge condition and a regulator for cutting off modes under the 
functional trace in \cite{Folkerts}. The literature which finds a non-zero quantum gravity 
correction to the running of gauge coupling are the following:
using loop-regularization in \cite{TangWu1}; in the presence of 
cosmological constant using gauge independent formulation 
of effective action (Vilkowisky-DeWitt technique) in \cite{Vilkovisky1984,TomsCosmo1, TomsCosmo2}
it was found that a nonzero contribution is achieved which is 
proportional to cosmological constant; using Vilkowisky-DeWitt technique
quadratic divergences were studied in \cite{TomsQuad1, TomsQuad2};
using functional renormalization group equation in \cite{Daum}.
In \cite{Robinson, Pietrykowski, Toms:2007, Rodigast2008, TangWu1, 
TomsCosmo1, TomsCosmo2, TomsQuad1, TomsQuad2} the study of the charge renormalization 
was done in the context of EH gravity in the spirit of effective field theory \cite{Donoghue1, Donoghue2, Donoghue3}, 
as the coupled system is non-renormalizable. Any results obtained from 
this can only be trusted at low energies as EH gravity is a low energy
limit of any fundamental theory of quantum gravity. In \cite{Daum, Folkerts} 
however Functional renormalization group has been used to study the problem 
in the spirit of asymptotic safety scenario \cite{Weinberg, AS_rev1, AS_rev2, AS_rev3, AS_rev4}.

In all these cases however it is not possible to give a proper meaning 
to the quantum corrections to the couplings, as in these cases the theory 
is non-renormalizable and has quadratic divergences \cite{Donoghue4}. These ambiguities 
don't occur in systems which are free of quadratic divergences and are
renormalizable. One such system is fourth order higher-derivative gravity
which is renormalizable to all loops \cite{Stelle}, and has recently been 
shown to be unitary \cite{NarainA1, NarainA2}. 
The first signature which provided the motivation for studying higher-derivative 
gravity came when quantum matter fields were studied on a curved background.
It was realized that at one-loop four kind of divergences appear 
\cite{Utiyama1962}: $\sqrt{-g}$, $\sqrt{-g} R$, $\sqrt{-g} R_\mn R^\mn$ 
and $\sqrt{-g} R^2$, where $R_\mn$ is the Ricci tensor of the background 
metric and $R$ is the corresponding Ricci scalar. This not 
only showed that the coupled system is non-renormalizable but also gave 
an important hint that perhaps considering a quantum theory 
of fourth order higher-derivative gravity coupled with matter will be more reasonable 
system to study as it may turn out to be completely renormalizable to all loops.
However historically things went around a slightly different direction.
A quantum theory of Einstein-Hilbert gravity was first studied.
This was also necessary, after all its a theory of classical gravity and 
describes a large number of phenomenas very accurately.
It was found in an one-loop study that the quantum theory of pure EH 
gravity is renormalizable on-shell \cite{thooft1}. In the same
paper it was also noticed that the theory is non-renormalizable 
even at one-loop when matter is included. This was further 
confirmed in \cite{Deser1, Deser2, Deser3, Deser4, Deser5, Deser6}.
This was somewhat expected as the gravitational coupling parameter,
Newton's constant $G$ has negative mass dimensions, 
but a thorough study was needed to explicitly check the 
renormalizability of the system. A two loop analysis 
of EH gravity assured the non-renormalizability of the 
theory \cite{Goroff1, Goroff2, vandeVen}. Such discouragements 
gave boost as to modify the gravity action in order to have a better ultraviolet behavior of theory.
Indeed it was found in \cite{Stelle} that when EH-gravity action is augmented
with fourth order higher-derivative terms, then the action is perturbatively
renormalizable to all loops in four spacetime dimensions. 
The higher derivative gravity action that was considered in \cite{Stelle}, 
in arbitrary space-time dimensions is given by,
\beq
\label{eq:hdg_act}
S_{\rm GR}
= \int \frac{ {\rm d}^d x \sqrt{-g}}{16 \pi G}
\biggl[
-R -\frac{1}{M^2} \left(
R_\mn R^\mn - \frac{d}{4(d-1)} R^2
\right) + \frac{(d-2) \omega}{4(d-1) M^2} R^2
\biggr] \, ,
\eeq
where $G$ is the Newton's constant, $M$ has dimensions of $mass$ while 
$\omega$ is dimensionless. Here the action is written in $d$-spacetime 
dimensions, for $2\leq d\leq4$. The most general fourth order derivative 
action that can be written includes term: Ricci scalar $R$, 
Weyl-square term $C_{\mn\rho\sg}C^{\mn\rho\sg}$,
Ricci scalar square term $R^2$ and Gauss-Bonnet term.
However for the dimensions under consideration, the 
Gauss-Bonnet term is topological and is not relevant in 
perturbative studies. This implies that 
one can re-express the $R_{\mn\rho\sg}R^{\mn\rho\sg}$
as a combination of $R_\mn R^\mn$ and $R^2$, thereby allowing 
one to rewrite $C_{\mn\rho\sg}C^{\mn\rho\sg}$ as a combination 
of $R_\mn R^\mn$ and $R^2$ (modulo factor of spacetime 
dimensions $d$), which is the second term in eq. (\ref{eq:hdg_act}).
The $d$-dependent coefficient in front of $R^2$ term is chosen
for later convenience, as in the propagator it produces pole
at mass $M/\sqrt{\omega}$. 

The action given in eq. (\ref{eq:hdg_act}), although is renormalizable in 
four spacetime dimensions to all loops, but suffers from a serious problem of unitarity
\cite{Stelle}. This can be seen by writing the propagator of the theory. In the Landau
gauge the metric propagator of the theory consist of three kind of terms: there is a 
massive spin-2 propagation with a pole at mass $M$;
there is a massive scalar propagation with a pole at mass $M/\sqrt{\omega}$;
and the usual massless propagator for the graviton. The massive spin-2 
propagator has five degrees of freedom and is called $M$-mode, while the 
massive scalar propagation has one degree of freedom and is called `Riccion'.
The massless graviton has two degrees of freedom. 
The propagator of the $M$-mode has a negative residue and is responsible 
for breaking the unitarity of theory \cite{Julve, Salam}.

Recently, higher-derivative gravity has been investigated in four dimensions \cite{NarainA1, NarainA2}. 
There using the one-loop beta function of the gravitational couplings, it is shown 
that in a certain domain of coupling parameters space the mass of the spin-2 mode 
(which has negative norm) runs in such a way so that it is always above the 
energy scale, as a result the propagator of $M$-mode never witnesses the pole.
This has useful consequences: one, there is never enough energy to create
this particle, so it never goes on-shell; second when it appears off-shell
it doesn't contribute to the imaginary part of scattering amplitude
(Cutkosky cut). In this way the theory remains unitary once the quantum 
corrections are taken into account, a feature which is absent in classical 
theory and at tree level. Furthermore, the authors of \cite{NarainA1, NarainA2}
also found that $G$ remains small for all energies and hence in the perturbative domain.
It vanishes at some finite energy albeit larger than the Planck energy. 
Therefore in this context the question of charge renormalization 
of gauge theories is raised within the perturbation theory to all 
orders in the Feynman loop expansion. The coupled system of 
higher-derivative gravity with the gauge fields is also renormalizable to
all loops \cite{Fradkin, Moriya}. This being free of quadratic divergences 
evades the criticism raised in \cite{Donoghue4}.
Recently gauge fields coupled 
with higher-derivative gravity has been studied in \cite{NAgauge}, where 
it was found that the leading contribution to the gauge coupling beta function 
in Feynman perturbation theory comes entirely due to quantum gravity 
effects and is seen to vanish to all loops. In \cite{NAgauge} we wrote that the 
proof holds only on-shell, however the proof is valid even off-shell.
Here in this paper we study the coupled system of gauge field with 
higher-derivative gravity in more detail. We study the unitarity of the 
coupled system and investigate the vanishing of the leading quantum 
gravity contribution in the gauge beta function to all loops off-shell. 

In section \ref{gaugest} we discuss the generic structure of the gauge 
coupling beta function for any gauge field coupled with gravity.
In section \ref{ea}, we build up the formalism and do a one-loop computation
for the quantum gravity correction to the gauge beta function. 
Here we show that `$a$' term is zero to one-loop in higher derivative gravity,
thereby computing the finite terms at one-loop for the abelian gauge fields. 
In section \ref{gravbeta}, we study the gauge contribution to the beta function of 
gravitational couplings and investigate the behavior of running gravitational couplings and 
unitarity of the coupled system. 
In section \ref{duality} we use symmetry arguments to study the `$a$' term of the 
gauge beta function to all loops. In section \ref{arbitGR}, we do one-loop computation 
to compute the quantum gravity contribution to gauge beta  function for 
arbitrary gravity action. We conclude in section \ref{conc}, with a discussion and 
implication of results. 

\section{Gauge Beta Function Structure}
\label{gaugest}

Generically, in Feynman path-integral study of gauge field coupled 
with gravity, the beta function of the gauge coupling in perturbation theory 
is lead by a term called ``$a$'' to all loops. This is the dominant 
contribution for small gauge couplings and is universally the same 
for all gauge theories coupled with gravity, including abelian gauge theories,
but depends on parameters present in pure gravity sector (in the 
presence of matter fields, this term also depends upon the  
information content of the matter sector), but it is independent of the 
gauge coupling. Being the dominant term for small gauge coupling, 
it can potentially overpower the 
asymptotic freedom of non-abelian gauge couplings, as was also demonstrated
in  \cite{Robinson, TomsQuad1, TomsQuad2}. In the following we define the `$a$' term more 
precisely, and show in section \ref{duality} that it vanishes to all loops.

\begin{figure}
\centerline{
\vspace{0pt}
\centering
\includegraphics[width=6.0in]{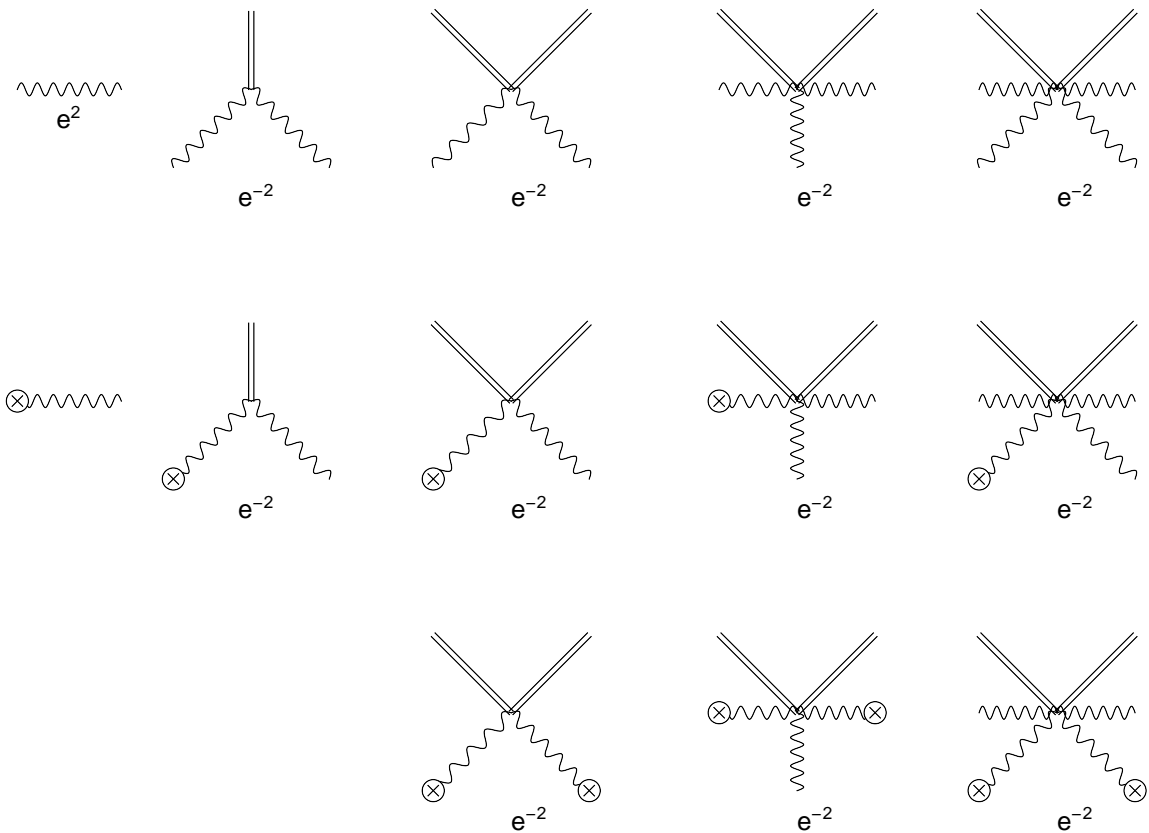}
}
 \caption[]{
Propagator of gauge field and vertices involving interaction of 
gauge field (wavy line) with the metric fluctuations (double line). 
The propagator (first graph) goes like $e^2$, while the 
vertices goes like $1/e^2$.
}
\label{fig:vertex}
\vspace{-5mm}
\end{figure}

The non-abelian gauge field action is given by, 
\beq
\label{eq:gaugeact}
S_{\rm gauge}=
- \frac{1}{4e^2} \int {\rm d}^4x \sqrt{-g} 
g^{\mu\al} g^{\nu\bt} F_{\mn}^a F_{\al\bt}^{a} \, ,
\eeq
where $e$ is the $SU(N)$-gauge coupling, $F_{\mu\nu}^a
=\partial_{\mu} A_{\nu}^a - \partial_{\nu}A_{\mu}^a
+ f^{abc}A_{\mu}^bA_{\nu}^c$, $A_{\mu}^a$ is the 
gauge vector potential and $f^{abc}$ are the structure constants. 
Defining quantum theory using the 
Feynman path-integral for the coupled action of gravity and 
gauge field given by eq. (\ref{eq:hdg_act} and \ref{eq:gaugeact}), we
use background field method to do the gauge fixing. In our 
case where the gravity action is renormalizable and unitary,
such a background gauge fixing guarantees a gauge invariant effective action
in dimensional regularization scheme \cite{thooft2, DeWitt1, DeWitt2, DeWitt3, Abbott}.
The running of gauge coupling constant satisfies the following 
generic equation,
\beq
\label{eq:beta_gauge_gr}
\frac{{\rm d}}{{\rm d} t} \left(
\frac{1}{e^2} \right) = \frac{a(M^2G, \omega, \cdots)}{e^2} 
+ b(e^2, M^2G, \omega, \cdots) \, ,
\eeq
where $t= \ln (\mu/\mu_0)$, the function `$a$' is independent of 
$e^2$ (the gauge coupling), but depends on the other couplings of the 
action (gravitational and matter sector couplings). It dependence 
on the gauge group comes implicitly through the dependence 
of other coupling on the number of generators of the group.
The `$b$'-term on the other hand depends on all the couplings 
present in the theory and the gauge group. The 
`dots' in the definition of `$a$' and `$b$' indicate the 
dependence on the matter couplings.
For small coupling this definition of the beta function is 
particularly useful as by construction $b$ is a regular function 
of $e^2$ at $e^2=0$ in Feynman perturbation theory.

\begin{figure}
\centerline{
\vspace{0pt}
\centering
\includegraphics[width=6.0in]{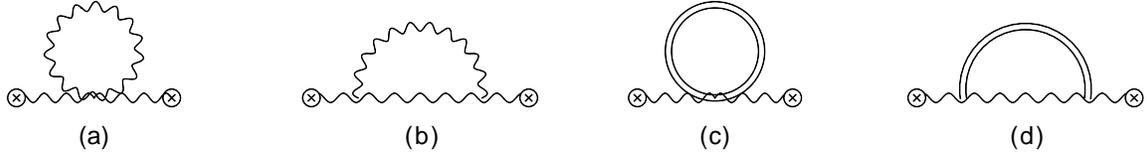}
}
 \caption[]{
One loop contribution to the running of gauge coupling.
The Feynman diagrams (a) and (b) represent the contribution
to the running of gauge coupling due to self interaction of the 
gauge fields (only for non-abelian gauge theories). The 
diagrams (c) and (d) are one-loop quantum gravity contribution to the 
gauge coupling (for all gauge theories). Here 
black dots represent the bare vertices, while wavy 
lines attached to circle with a cross represent external legs.
  }
\label{fig:1loop}
\vspace{-5mm}
\end{figure}

From the gauge field action given in eq. (\ref{eq:gaugeact}), 
we can apply the background field formalism to obtain the 
propagator of the fluctuating gauge field and the various vertices 
involving interaction of metric fluctuation with both the background 
gauge field and the gauge field fluctuation (written in appendix \ref{gaugeverprop}).
As this formalism is explicitly gauge invariant in the 
background field by construction, therefore the gauge coupling
and the wave-function renormalization are the same in this approach. 
This also means that in this picture one would only have to consider 
the two-point green function of the background field \cite{DeWitt3, Abbott}.
We therefore consider vertices with at most two background gauge 
field lines. The set of vertices and propagator for the gauge field 
are depicted in the Fig. \ref{fig:vertex}. Counting the powers 
of $e^2$, it is easy to realize that the propagator 
for the gauge fluctuation field is proportional to $e^2$,
while the background gauge field line carries no power of $e^2$.
All the vertices are proportional to $1/e^2$. From the vertices in Fig. \ref{fig:vertex}, 
we note that any vertex which involve only two gauge field line 
(either background or fluctuation field) depicts interactions where gravity couples 
with gauge field purely due to its energy, while any vertex involving 
more than two gauge field line (either background or fluctuation field) are interaction 
where there is also charge interactions.

In this formalism at one-loop there are four diagrams that give contribution to the running 
of gauge coupling: the first two diagrams come because of self interaction 
between gluons and are only present for non-abelian gauge fields, 
the other two diagrams give quantum gravity contribution to the 
running gauge coupling, as shown in Fig. \ref{fig:1loop}. 
The first two diagrams are only present for non-abelian gauge 
fields, while the other two diagrams are present for all gauge 
fields. Studying the $e^2$ dependence of the diagrams we notice 
that both the one-loop diagrams that give quantum gravity contribution 
to the running of gauge coupling are proportional to $1/e^2$, 
thereby contributing to `$a$' term alone in eq. (\ref{eq:beta_gauge_gr}).  

\begin{figure}
\centerline{
\vspace{0pt}
\centering
\includegraphics[width=5in]{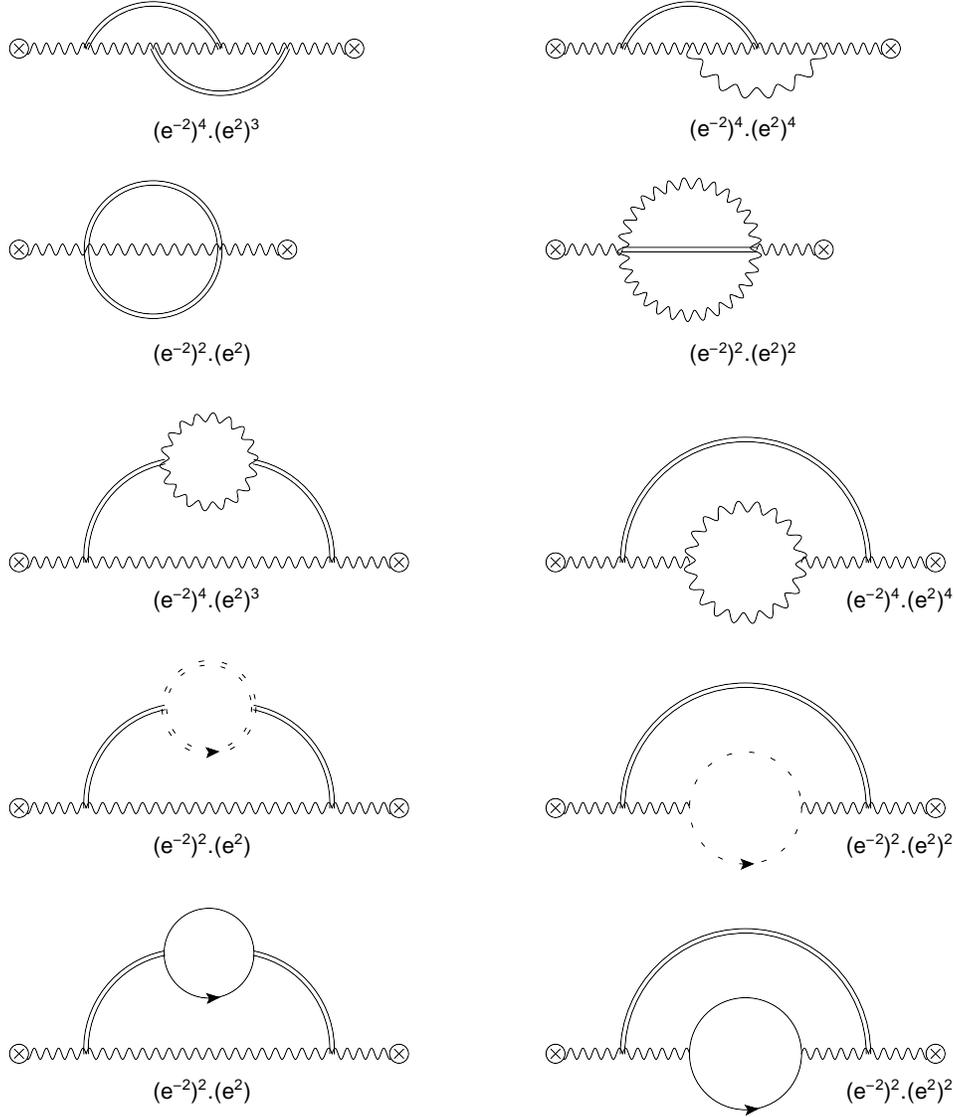}
}
 \caption[]{
Various Feynman diagrams in the perturbation theory of gauge field 
coupled with gravity. First column contains diagrams contributing 
to `$a$'-term, while second column contain diagrams 
contributing to `$b$'-term.
All diagrams containing 3- or 4-gluon vertex 
give contributions to `$b$'-term of the beta function. Matter 
loops attached to gluon line contributes to `$b$-term, while the 
one attached to the graviton line contributes to `$a$'-term.
There are two kinds of ghost: gravitational ghost (which interact 
with only background and fluctuating metric field) and gauge field
ghosts (which interact with background metric, background and 
fluctuating gauge field line). Gravity ghost contribute to 
`$a$'-term, while gauge field ghost contribute to `$b$'-term.
}
\label{fig:AandB}
\end{figure}

Now we study the nature of `$a$'-term to all loops. Here the task is simplified
by being in background field formalism where one only has to consider
two point correlation function of the background gauge field. We notice
that any diagram involving a vertex with three- or four-gluon exchange  
will only contribute to the `$b$'-term of the gauge beta function.
This is easily verified by counting the powers of $e^2$. For example
an inclusion of 3- or 4-gluon vertex in any diagram 
contributing to the `$a$'-term (meaning that it goes like $1/e^2$),
will result in an increase in the power of $e^2$ by at least one factor 
of $e^2$, meaning such diagram will eventually contribute to `$b$'-term.
This observations can be more clearly checked 
from the Fig. \ref{fig:AandB}, where appearance of 3- or 4- gluon 
vertex in a diagram shows that it gives contribution to `$b$'-term.  
So effectively one can ignore these kind of diagrams, this 
is equivalent to ignoring the three and four gluon terms in the 
expansion of $F^a_\mn F^{a\mn}$, which is like considering 
$N^2-1$ abelian gauge fields. Diagrams arising from these terms 
will necessarily contribute to `$a$' term alone, and will be same 
that is obtained in $U(1)$ gauge field case.

Regarding matter it should be noticed that their interaction with the 
gauge fields (both background and fluctuation) arise via
the kinetic term of the matter fields, where gauge covariant derivative 
is introduced in order to preserve the gauge invariance of the 
matter action. Such interaction terms don't carry any power of 
$e^2$. Matter fields also interact with metric fluctuation. Due to 
non-linearity of the gravity, such interaction terms are infinite in number.
Matter field contribute to the gauge beta function through the loops.
When matter field loop is attached to the gauge field line, then 
it contributes to the `$b$'-term, however when it is attached to the
internal graviton line, then it contributes to the `$a$'-term. For example
any diagram contributing to the `$a$'-term will still contribute to 
`$a$'-term when matter loop is attached or inserted in a graviton line.
This is because vertices involving interaction of matter with graviton 
don't carry any power of $e^2$. But on other hand when matter loop 
is inserted/attached to gluon line then the power of $e^2$ increase 
by one. This is because although the new vertices added don't carry 
any power of $e^2$, but insertion of such loops splits the 
gluon propagator in two, thereby increasing the power of $e^2$
by one. Such modified diagrams will then contribute to `$b$'-term. 
This can be further verified by examining the two-loop graphs 
given in Fig. \ref{fig:AandB}.

The case of ghost field is a bit different. There are two kind of ghost:
gravitational and gauge ghosts. These two ghosts are not independent 
and mix with each other. This is due to the fact that the gauge field undergoes 
both deffeomorphism and gauge transformation. This mixing looks 
problematic but is actually innocuous, as it never contributes
to any feynman diagram. This will become more clear later 
in the section of gauge fixing and ghosts \ref{gfghost}. As a result 
the gravity and gauge field ghost become independent. 
Gravity ghosts only interact either with the background metric 
or with the metric fluctuation field, as a result of which their presence 
doesn't alter the power of $e^2$ of the diagram. Gauge field ghost 
on the other hand interacts with the background metric, background
gauge field and the gauge fluctuation field. All these interactions don't
carry any powers of $e^2$. This means that when ghost loops are inserted 
in the gauge field line of the diagrams, they tend to increase the power of
$e^2$ by one. Thus any diagram which was originally contributing to `$a$'-term
alone, after the inclusion of the gauge field ghost loop will 
contribute to `$b$'-term. This can be further verified by 
examining the two loop graphs involving the ghost loops depicted 
in Fig. \ref{fig:AandB}.

These arguments tell that gauge field ghosts when present in the 
diagrams will always contribute to the `$b$'-term of the gauge beta
function. On the other hand in the case of matter fields, when the loop
is attached to the graviton line then it contributes to `$a$'-term, while 
when it is attached to the gluon line, it contributes to the `$b$'-term. 
Focusing just on the diagrams contributing to 
`$a$'-term alone, is equivalent to considering parts of matter action which
doesn't depend on the gauge field. This is same as saying that 
one is considering uncharged fields. Thus to sum up, the kind of terms
of the total matter action (gauge field action plus the matter action) 
that contribute to the `$a$'-term alone to all loops 
are: for gauge field action, terms which 
don't have 3- or 4-gluon interactions; for the matter action, terms
which are independent of the gauge field (same as considering uncharged
field). In Fig. \ref{fig:AandB}, although we have demonstrated only up to
two loops, but this argument is valid to all loops as discussed in previous 
paragraphs. However `$a$' term is independent of the gauge coupling $e^2$,
but can depend upon the gauge group (number of generators of gauge 
group) via the gravitational and matter couplings dependence 
on the gauge group. The universality is a manifestation of the fact that the metric fluctuations
interact universally to all gauge fields via its energy. 

The formal solution of eq. (\ref{eq:beta_gauge_gr}) can be written as,
\beq
\label{eg:e^2run}
\frac{1}{e^2} = e^{ \int_0^t a \, {\rm d} t^{\prime}} \biggl(
\frac{1}{e_0^2} + \int_0^t {\rm d}t^{\prime} 
\, b \, e^{- \int_0^{t^{\prime}} a \, {\rm d} t^{\prime\prime}}
\biggr)
\eeq
It is evident that for small $e^2$ the running of $e^2$ depends 
more dramatically on the sign of $a$. If $a$ is negative then 
$e^2$ diverges as $e^{|a| t}$ for large $t$. 
If $a$ is positive, then for large $t$, $e^2$ vanishes 
faster than $e^{-|a| t}$ as considered in \cite{Robinson, TomsQuad1, TomsQuad2}.
If $a=0$, then the standard behavior of the
running of gauge coupling qualitatively holds. The above equation is valid to any order in the loop 
expansion. So the qualitative behavior of $e^2$ namely asymptotic freedom 
can remain unaltered if $a=0$ to all loops. 
This was examined and studied in \cite{NAgauge} and found that 
$a=0$ to all loops off-shell (mistakenly written on-shell,
but proof is valid off-shell), while explicitly showing that it is zero 
at one-loop off-shell (independent of gravity action, regularization scheme
and gauge fixing condition). Here we study the coupled system in 
more detail addressing the `$a$'-term to all loops off-shell and 
will also study the unitarity of the coupled system.

\section{Effective Action}
\label{ea}

In this section we study the coupling of higher derivative gravity with gauge field, 
to compute the quantum gravity correction to the running of gauge couplings.
To begin with we consider the path-integral of the coupled gravity and gauge
action given in eq. (\ref{eq:hdg_act} and \ref{eq:gaugeact}) respectively. 
This is given by,
\begin{equation}
\label{eq:pathint}
Z
= \int \, {\cal D} \g_\mn
{\cal D} A_{\mu}^a 
\exp \biggl[
i\biggl(
S_{\rm GR} + S_{\rm gauge} 
\biggr)
\biggr] \, ,
\end{equation}
where $S_{\rm GR}$ is the higher derivative gravity action
given in eq. (\ref{eq:hdg_act}), $S_{\rm gauge}$ is the 
gauge field action given in eq. (\ref{eq:gaugeact}).

We study the deffeomorphism invariant gravity action and gauge 
invariant gauge field action using background field method \cite{DeWitt3, Abbott}.
This has an advantage as by construction it preserves the background gauge field 
invariance of the effective action. In this formalism the 
quantum fields of gravity and gauge theory are decomposed
into a background and a fluctuation. Keeping the background fixed,
the invariance of the fluctuation field of the full action is broken by
constraining the fluctuation fields. This procedure results in the 
generation of certain auxiliary fields called ghosts. The overall
resulting action after applying the constraint in the presence of ghost
still possess invariance over the background metric and the background gauge field,
thereby producing a background gauge invariant effective action.

Writing the quantum metric $\g_\mn$ as $\bg_\mn+h_\mn$, where 
$\bg_\mn$ is some arbitrary fixed background and $h_\mn$ is the 
metric fluctuation, we expand the full action in powers of $h_\mn$.
As the background metric is fixed, therefore the path-integration measure
over the quantum metric $\g_\mn$ gets replaced by the measure 
over the fluctuation field $h_\mn$. Integrating over the fluctuation 
field $h_\mn$ completely also implies that they will only appear 
as virtual particles inside the loop and never as external legs.
Effective action computed after the path-integration over the 
fluctuation field $h_\mn$ being completely deffeomorphism invariant 
in the background metric, allows one to choose a particular background 
metric in order to simplify the computation. In particular choosing
$\bg_\mn = \eta_\mn+H_\mn$ (while still keeping $H_\mn$ generic), 
allows one to use the methodology and formalism of the flat 
spacetime quantum field theory. In this way of working it is possible 
to attribute particle notions to the fields $H_\mn$ and $h_\mn$, where
$h_\mn$ will behave as a virtual particle, while $H_\mn$ will 
act as a external particle corresponding to $h_\mn$, exactly as is the 
case in flat spacetime QFT for usual scalar or spinor fields. 
Attributing particle notions to the fluctuation field $h_\mn$ and $H_\mn$ 
allows one to also consider scattering matrix amplitudes. 
Looking from a different angle it is quickly realized that 
expanding the gravity action around a flat spacetime and 
calling the perturbations around it to be $h_\mn^{\prime}$, 
one obtains a highly nonlinear gauge theory in the field 
$h_\mn^{\prime}$. Treating this gauge theory along the lines 
of background field method, where now the quantum field 
$h_\mn^{\prime}$ is written as $H_\mn+h_\mn$, allows one to 
quickly see that this is a flat spacetime QFT with $H_\mn$ 
as an external leg and $h_\mn$ as the internal line. Integrating
over the fluctuation field $h_\mn$, gives one an effective action 
as a functional of $H_\mn$ field, which is the effective action 
for an arbitrary background expanded about flat spacetime.

In this language one can setup Feynman perturbation theory
by expanding the original action of gravity and matter 
in powers of $h_\mn$ and $H_\mn$. This series of terms
will contain propagator of the $h_\mn$ field and various vertices
involving interactions of external field $H_\mn$ with internal 
field line $h_\mn$. This is the expanded bare action of the theory and
carries infinite number of vertices due to nonlinear nature of 
gravitational field. However, as we still have background gauge 
invariance, therefore one only needs to do the expansion of the bare action
only up to certain finite order in $H_\mn$. This means that for 
studying the behavior of term linear in Ricci scalar curvature under quantum 
corrections, it is sufficient to expand the bare action up to linear order in $H_\mn$, and for 
studying the behavior of terms quadratic in curvature it is 
sufficient to expand the bare action up to quadratic in $H_\mn$. 
Such expansion provide all the vertices that will be relevant for 
study of these kind of terms. This is the privilege of background field method. 

For a one-loop computation on the other hand it is sufficient to expand
the bare action up to second order in $h_\mn$. This although finite 
in the order of $h_\mn$, but is infinite in order of $H_\mn$. The restriction on the 
order of $H_\mn$ comes from the requirement of the kind of term that 
is to be investigated in the effective action. 
For example probing the issues related to cosmological constant, it is 
sufficient to put $H_\mn=0$ {\it i.e.} truncating the series 
at zeroth order in $H_\mn$. Similarly exploring the Ricci scalar 
term of the effective action demands to consider term up to linear 
in $H_\mn$ and so forth. Investigating quantum gravity contribution to 
charge renormalization will demand accordingly to consider 
terms up to zeroth order in $H_\mn$, {\it i.e.} putting $H_\mn=0$.

\subsection{Gauge Fixing and Ghosts}
\label{gfghost}

The path integration over the gauge and gravitational fields 
is not well defined. This is due to fact that for certain field 
configurations the integrand of the functional integral is unity,
thereby implying a diverging path integral.
This is because gauge and deffeomorphism transformation allows 
to choose field configurations for which the action is zero. 
Besides this, the measure of the functional integral over the 
gauge field is also ill defined, due to over counting of 
gauge orbits. Both these problems are evaded by
constraining both the gauge and metric field thereby breaking the
gauge and deffeomorphism invariance of the path integral. 
This procedure of breaking the invariance of the path integral 
is echoed by giving rise to ghosts, which is elegantly 
studied through the methodology of Faddeev-Popov
 \cite{Faddeev}.

However, in this style of obtaining the effective action 
it is difficult to see the invariance, once path integration is
performed. This is overcome by using 
background field method, which explicitly assures the 
invariance of effective action in the background fields.
In this picture the path integral is over the fluctuation 
fields $h_\mn$ and ${\cal A}_\mu^a$. Gauge fixing of the
path integral is done in such a way so that it breaks the 
invariance over the fluctuation fields, while still preserving 
the residual invariance over the background fields. In the 
following we will discuss in detail how the background 
gauge fixing is done for this coupled system of gauge 
and gravity fields.

The coupled action of gauge and gravity field being 
deffeomorphism invariant in the field variables implies,
that for an arbitrary vector field $\ep^\rho$, the action
should be invariant under the following transformation 
of the metric field variable,
\beq
\label{eq:gaugetrgamma}
\de_D \g_\mn = {\cal L}_\ep \g_\mn = \ep^\rho \pt_\rho \g_\mn + \g_{\mu\rho} \pt_\nu
\ep^\rho + \g_{\nu\rho}\pt_\mu \ep^\rho \, ,
\eeq
where ${\cal L}_\ep \g_\mn$ is the Lie derivative of the quantum metric 
$\g_\mn$ along the vector field $\ep^\rho$. This metric field when 
decomposed in to a background ($\bg_\mn$) and fluctuation ($h_\mn$), 
allows one to obtain the transformation of the tensor field $h_\mn$
while keeping the background fixed. This will imply the following 
transformation of $h_\mn$.
\beq
\label{eq:gaugetrh}
\de_D h_\mn = \bar{\nabla}_\mu \ep_\nu + \bar{\nabla}_\nu \ep_\mu
+ \ep^\rho \bar{\nabla}_\rho h_\mn + h_{\mu\rho} \bar{\nabla}_\nu \ep^\rho
+ h_{\nu\rho} \bar{\nabla}_\mu \ep^\rho \, ,
\eeq
where $\bar{\nabla}$ is the covariant derivative whose connection is
constructed using the background metric. This is the full 
transformation of the tensor field $h_\mn$. Ignoring terms 
which are linear in $h_\mn$ allows one to investigate 
only one-loop effects, while they are kept in the complete 
analysis involving higher-loop studies.

The gauge field action given in eq. (\ref{eq:gaugeact})
on the other hand has two kind of invariances: deffeomorphism 
invariance and local gauge invariance. For an arbitrary vector field
$\ep^\rho$ and a color vector field $\lam^a$, gauge field 
action is invariant under the following transformation of the
gauge field $A_\mu^a$.
\bea
\label{eq:gaugeTR}
{\rm Diff:} \, \, \Rightarrow \,\, 
&& 
\de_{D} A_\mu^a = {\cal L}_{\ep} A_\mu^a = \ep^\rho \pt_\rho A_\mu^a 
+ A^a_\rho \pt_\mu \ep^\rho \, ,
\notag \\
{\rm Gauge:} \, \, \Rightarrow \,\, 
&&
\de_g A_\mu^a = \pt_\mu \lam^a - f^{abc} \lam^b A_\mu^c \, .
\eea
Decomposing the gauge field in to a background $\bar{A}_\mu^a$ 
and a fluctuation ${\cal A}_\mu^a$, implies the following transformation
of the fluctuating field ${\cal A}_\mu^a$ while keeping the 
background fixed.
\beq
\label{eq:gaugetrA}
\de {\cal A}_\mu^a = 
\ep^{\rho} \bar{\nabla}_\rho \bar{A}_\mu^a
+ \bar{A}_\rho^a \bar{\nabla}_\mu \ep^\rho 
+ \ep^{\rho} \bar{\nabla}_\rho {\cal A}_\mu^a
+ {\cal A}_\rho^a \bar{\nabla}_\mu \ep^\rho 
+ \bar{\nabla}_\mu \lam^a
+ f^{abc} \bar{A}_\mu^b \lam^c
+ f^{abc} {\cal A}_\mu^a \lam^c \, ,
\eeq
where the first four terms on the \textit{rhs} of the equality 
denote the transformation of the fluctuation field ${\cal A}_\mu^a$
under the deffeomorphism, while the last three terms denote 
the transformation under the $SU(N)$ gauge group
(for abelian gauge fields, the terms proportional to 
structure constants will be absent). This is the complete 
transformation of the fluctuation field ${\cal A}_\mu^a$, under 
which the gauge field action will be invariant. 
This invariance of the action under the transformation of the 
fluctuation field is broken by choosing an appropriate 
gauge fixing condition for breaking the deffoemorphism and gauge 
invariance of the action. Choosing a gauge fixing action which 
is invariant under the background gauge transformation but breaks the
invariance under the transformation of the fluctuation field, further
assures the preservation of background gauge invariance of the effective action.

The gauge fixing action chosen for fixing the invariance under the 
transformation of the metric fluctuation field is given by,
\beq
\label{eq:gravity_GF}
S^{\rm gravity}_{GF} = \frac{1}{32 \pi G \alpha} 
\int {\rm d}^dx \sqrt{-\bg} \biggl(\bnb^{\rho} h_{\rho\mu}
- \frac{1+\rho}{d} \bnb_\mu h \biggr) \, Y^\mn \, 
\biggl(\bnb^{\sigma}h_{\sigma\nu} 
-\frac{1+\rho}{d} \bnb_\nu h\biggr) \, .
\eeq
where $\al$ and $\rho$ are gauge parameters, while $Y_\mn$
is either a constant or some differential operator depending 
upon the gravity theory under consideration. In the case of 
higher-derivative gravity like the one described by action in
eq. (\ref{eq:hdg_act}), we consider higher-derivative type
gauge fixing by taking $Y_\mn = (-\bg_\mn \bar{\Box}
+ \bt \bnb_\mu \bnb_\nu)$, where $\Box = \bnb_\mu \bnb^\mu$.
Similarly the fluctuating gauge field is constrained by 
choosing the following gauge fixing action,
\beq
\label{eq:sgf_gauge}
S^{\rm gauge}_{GF} = \frac{1}{2\xi} \left( -\frac{1}{e^2} \right)
\int {\rm d}^dx \sqrt{-\bg} \left( \textbf{D}_{\mu} \mathcal{A}^{a\mu} \right)^2 \, ,
\eeq
where $\xi$ is a gauge parameter.
Here $\textbf{D}$ is the covariant derivative constructed 
with background metric and the background gauge field. 
Its action on the fluctuation gauge field is given by,
\beq
\label{eq:covantDer}
\textbf{D}_\mu {\cal A}_\nu^a = \pt_\mu {\cal A}_\nu^a
- \bar{\G}_\mu{}^\al{}_{\nu} {\cal A}_\al^a + f^{abc} \bar{A}_\mu^b {\cal A}_\nu^c \, .
\eeq

The ghost action for these gauge fixing conditions can be obtained 
by using the Faddeev-Popov trick \cite{Faddeev}. 
Calling in general the gauge fixing condition for 
gravitational field $h_\mn$ to be $F_{1\mu}=0$, while 
the gauge fixing condition for the fluctuating gauge field
${\cal A}_\mu^a$ to be $F_{2}^a =0$ (in the present case
we have $F_{1\mu} = \bnb^{\rho} h_{\rho\mu}
- \frac{1+\rho}{d} \bnb_\mu h $ and 
$F_2^a = \textbf{D}_{\mu} \mathcal{A}^{a\mu}$), 
we introduce them in the path integral of the fluctuating fields
by multiplying it with unity in the following form.
\beq
\label{eq:pathunity}
1= \int {\cal D} F^{\ep\lam}_{1\mu} {\cal D} F_2^{a\ep\lam} \left(\det Y \right)^{\frac{1}{2}}
\exp \biggl[ \frac{i}{32 \pi G \al} \int {\rm d}^dx \sqrt{-\bg} F^{\ep\lam}_{1\mu}
Y^\mn F^{\ep\lam}_{1\nu} + \frac{i}{2 \xi e^2} \int {\rm d}^dx \sqrt{-\bg}
F_2^{a\ep\lam} F^{\ep\lam}_{2a} \biggr] \, ,
\eeq
where $F_{1\mu}^{\ep\lam}$ and $F_2^{a\ep\lam}$ are the gauge transformed 
$F_{1\mu}$ and $F_2^a$ respectively. If $Y^\mn$ did not contain any derivatives, then this
determinant would be trivial, however it will not be so if 
$Y^\mn$ contains derivative operators, which is the case 
in higher-derivative gravity. 
The original path integral (without gauge fixing) being invariant 
under transformation eq. (\ref{eq:gaugetrh} and \ref{eq:gaugetrA})
of the fluctuation fields $h_\mn$ and ${\cal A}_\mu^a$, implies
that the integration variable $h_\mn$ and ${\cal A}_\mu^a$ can be
changed to transformed fields $h_\mn^\ep$ and ${\cal A}_\mu^{a\ep\lam}$.
This transformation don't give rise to any non-trivial 
jacobians in the path-integral measure of the 
gauge and gravity field. Writing the measure over 
$F^{\ep\lam}_{1\mu}$ and $F_2^{a\ep\lam}$, as measure over the 
transformation variables $\ep^\rho$ and $\lam^a$, introduces
a non-trivial jacobian in the path integral. This is the 
Faddeev-Popov determinant and is worked out as follows.
\bea
\label{eq:gaugejacob1}
&&
dF_{1\mu}^{\ep\lam} = \frac{\pt F_{1\mu}}{\pt \ep^\rho} d\ep^\rho
+ \frac{\pt F_{1\mu}}{\pt \lam^b} d \lam^b \, , \notag \\
&&
d F_2^{a\ep\lam} = \frac{\pt F_2^a}{\pt \ep^\rho} d \ep^\rho
+ \frac{\pt F_2^a}{\pt  \lam^b} d \lam^b \, .
\eea
This means that the measure ${\cal D} F_{1\mu}^{\ep\lam} {\cal D} F_2^{a\ep\lam}$
transforms as,
\beq
\label{eq:gaugejacob2}
{\cal D} F_{1\mu}^{\ep\lam} {\cal D} F_2^{a\ep\lam}
= \det \left(
\begin{large}
\begin{array}{c c}
\frac{\pt F_{1\mu}}{\pt \ep^\rho} & \frac{\pt F_{1\mu}}{\pt \lam^b} \\
 & \\
\frac{\pt F_2^a}{\pt \ep^\rho} & \frac{\pt F_2^a}{\pt  \lam^b}
\end{array}
\end{large}
\right) {\cal D} \ep^\rho {\cal D} \lam^b \, .
\eeq
In the background field formalism, this jacobian consist of background 
covariant derivative, background and fluctuation fields.
As the determinant is independent of the transformation 
parameters $\ep^\rho$ and $\lam^a$, therefore 
it can be taken out of the functional integral over $\ep^\rho$ and $\lam^a$.
Changing the integration variable from $h_\mn^\ep$ 
and ${\cal A}_\mu^{a\ep\lam}$ to $h_\mn$ and ${\cal A}_\mu^a$
respectively, and ignoring the infinite constant generated 
by integrating over $\ep^\rho$ and $\lam^a$, 
gives us the gauge fixed path integral including the determinants. 

These functional determinants can be exponentiated by making use 
of appropriate auxiliary fields. Writing the functional 
determinant $(\det Y)^{1/2}$ as a product of two 
determinants $(\det \, Y) \times (\det \, Y)^{-1/2}$, allows us 
to combine the former with the Faddeev-Popov determinant in
eq. (\ref{eq:gaugejacob2}), which is then exponentiated 
by making use of anti-commuting auxiliary fields, while the 
later determinant $(\det \, Y)^{-1/2}$ is exponentiated 
by making use of commuting auxiliary fields. The former 
auxiliary fields are known as Feddeev-Popov ghosts, while the
auxiliary field in the later case is known as 
Neilsen-Kallosh ghosts \cite{Kallosh, Nielsen}.
The path integral of the full ghost sector is given by,
\bea
\label{eq:ghostpath}
&&
\int {\cal D}\bar{C}_\mu {\cal D} C_\nu {\cal D} \bar{c}^a {\cal D} c^b {\cal D} \ta_\al
\,\,
\exp \biggl[
-i \int {\rm d}^dx \sqrt{-\bg} \, 
\left(
\begin{array}{c c}
\bar{C}_\mu & \bar{c}^a
\end{array}
\right) \cdot
\notag \\
&&
\left(
\begin{array}{c c}
Y^\mn \, \pt F_{1\nu} /\pt \ep^\rho & Y^\mn\pt F_{1\nu}/ \pt \lam^b \\
 & \\
\pt F_2^a / \pt \ep^\rho & \pt F_2^a/ \pt  \lam^b
\end{array}
\right) \cdot 
\left(
\begin{array}{c}
C^\rho \\
c^b
\end{array}
\right) 
-\frac{i}{2} \int {\rm d}^dx \sqrt{-\bg} \, \ta_\al Y^{\al\bt} \ta_\bt 
\biggr] \, ,
\eea
where $\bar{C}_\mu$ and $C_\nu$ are anti-commuting fields arising 
from the gauge fixing in the gravitational sector, while $\bar{c}^a$ and
$c^b$ are anti-commuting fields arising from the gauge fixing in the 
gauge field sector and $\ta_\mu$ is the commuting ghost 
arising due to fact that $Y_\mn$ contains derivatives.
It is crucial to note here that in the coupled system of gauge 
field with gravity, there will always be mixing between the 
gravitational and gauge ghosts. The mixing term
$Y^\mn\pt F_{1\nu}/ \pt \lam^b$ will arise, if we choose a 
gauge fixing condition for the gravitational field $h_\mn$ 
such that it also contain terms involving gauge fluctuation field
${\cal A}_\mu^a$. In the absence of such term, this mixing
term will be always zero. However, the mixing term
$\pt F_2^a / \pt \ep^\rho$ will be always non-zero, as the 
gauge fluctuation field ${\cal A}_\mu^a$ undergoes both 
deffeomorphism and gauge transformation.

In the case when $F_{1\mu}$ and $F_2^a$ are given as in
eq. (\ref{eq:gravity_GF}) and (\ref{eq:sgf_gauge}) respectively,
the Faddev-Popov ghost action is given by,
\beq
\label{eq:ghostact}
S^{FP}_{\rm gh} = - \int {\rm d}^dx \sqrt{-\bg}
\biggl[
\bar{C}_\mu X^\mu_{1\rho} C^\rho
+ \bar{c}^a X^a_{2\rho} C^\rho
+ \bar{c}^a X_3^{ab} c^b \, ,
\biggr] \, ,
\eeq
where,
\bea
\label{eq:ghostet}
X^\mu_{1\rho} &=& 
(\bg^\mn \bar{\Box} + \bt \bnb^\mu \bnb^\nu)
\biggl[
\bnb_\rho \bnb_\nu + \bg_{\nu\rho} \Box
- \frac{2(1+\rho)}{d} \bnb_\nu \bnb_\rho + \bnb_\rho h_{\sg\nu} \bnb^\sg
+ \bnb^\sg\bnb_\rho h_{\sg\nu} 
\notag \\
&&
+ \bnb^\sg h_{\nu\rho} \bnb_\sg + h_{\nu\rho}\bar{\Box}
+ \bnb^\sg h_{\sg\rho} \bnb_\nu 
+ h_{\sg\rho} \bnb^\sg \bnb_\nu
\notag \\
&&
- \frac{1+\rho}{d} \biggl(
\bnb_\rho h\bnb_\nu + \bnb_\nu\bnb_\rho h 
+ 2 \bnb_\nu h_{\sg\rho} \bnb^\sg
+ 2 h_{\sg\rho}\bnb_\nu \bnb^\sg
\biggr)
\biggr] \, ,
\\
X^a_{2\rho} &=& 
\bnb_\rho\bar{A}_\sg^a\bnb^\sg
+ \bnb^\sg\bnb_\rho \bar{A}_\sg^a
+ \bnb^\sg\bar{A}_\rho^a \bnb_\sg
+ \bar{A}^a\rho \bar{\Box}
+ \bnb_\rho {\cal A}^a_\sg \bnb^\sg
+\bnb_\sg\bnb_\rho{\cal A}^{a\sg}
+ \bnb^\sg {\cal A}^a_\rho \bnb_\sg
\notag \\
&&
+ {\cal A}^a_\rho \bar{\Box}
+ f^{abc} \bar{A}^{b\sg} \bnb \bar{A}^c_\sg
+ f^{abc} \bar{A}^{b\sg} \bar{A}^c_\rho\bnb_\sg
+ f^{abc} \bar{A}^{b\sg}\bnb_\rho {\cal A}^c_\sg
+ f^{abc} \bar{A}^{b\sg} {\cal A}^c_\rho \bnb_\sg \, ,
\\
X_3^{ab} &=& 
\de^{ab} \bar{\Box} 
- f^{abc} \bnb^\sg \bar{A}^c_\sg
- 2 f^{abc}\bar{A}^c_\sg \bnb^\sg
-f^{abc} \bnb^\sg {\cal A}^c_\sg
-f^{abc} {\cal A}^c_\sg \bnb^\sg
\notag \\
&&
+ f^{aec} f^{cdb} \bar{A}^{e\sg} \bar{A}^d_\sg
+ f^{aec} f^{cdb} \bar{A}^{e\sg} {\cal A}^d_\sg \, .
\eea
It should be noted that for the kind of gauge fixing condition 
considered here in eq. (\ref{eq:gravity_GF}) and (\ref{eq:sgf_gauge}), 
only the mixing term $\pt F_2^a / \pt \ep^\rho$ 
is non-zero, while the other mixing term $Y^\mn\pt F_{1\nu}/ \pt \lam^b$ is zero.
This means that there is no interaction between gravitational 
ghost $\bar{C}_\mu$ and gauge ghost $c^a$. This has an important
consequence. This will imply that in this kind of gauge fixing, there 
will not be any closed loop Feynman diagram with one line in the 
loop to be gravitational ghost, while the other gauge ghost. Therefore,
the presence of mixing term becomes completely innocuous. Had the 
other mixing term in the Faddeev-Popov determinant 
$Y^\mn\pt F_{1\nu}/ \pt \lam^b$ been nonzero, then we will have 
interaction term in the ghost action involving $\bar{C}_\mu$ and 
$c^a$. In such cases there will graphs involving loop with 
one gravitational ghost and one gauge ghost.

\subsection{Gravitational Field Propagator}
\label{propa}

The propagator of the fluctuating quantum field $h_\mn$ is given
is obtained by expanding the bare gravitational action eq. (\ref{eq:hdg_act}) 
up to second order in $h_\mn$ around flat spacetime background. 
This is given by,
\bea
&&
\label{eq:2ndVar_GR}
\de^2 S_{\rm GR} = \int \frac{ {\rm d}^d x}{32 \pi G}
\biggl[
-h \pt_\mu \pt_\nu h^\mn + \frac{1}{2} h\Box h
- \pt_\nu h^\mn \pt_\rho h^{\rho}_{\mu} - \frac{1}{2} h_\mn \Box h^\mn
\notag \\
&&
- \frac{1}{2M^2} (\pt_\rho \pt_\nu h^{\rho}_{\mu}
+\pt_\rho \pt_\mu h^{\rho}_{\nu} - \pt_\mu\pt_\nu h
- \Box h_\mn)(\pt^\sg \pt^\nu h_{\sg}^{\mu} 
+ \pt^\sg \pt^\mu h_{\sg}^{\nu} - \pt^{\mu} \pt^{\nu} h
-\Box h^{\mn})
\notag \\
&&
+  \frac{2}{M^2} \frac{d+(d-2) \omega}{4d(d-1)}
(\pt_{\mu} \pt_{\nu} h^\mn - \Box h) (\pt_\al \pt_\bt h^{\al\bt}
-\Box h)
\biggr] \, .
\eea
By making use of the complete set of orthogonal projectors in flat space
for the symmetric rank-2 tensor field (see appendix \ref{app_proj}), 
this can be re-written in momentum space in term of projectors,
\bea
\label{eq:2ndVar_GRproj}
\de^2 S_{\rm GR} = \frac{1}{32 \pi G} \int \frac{{\rm d}^dq}{\pd}
h_{\mn} \biggl[
-\frac{q^2}{2M^2} (q^2 - M^2) P_2^{\mn \rho\sg}
+ \frac{(d-2)\omega}{2M^2}  q^2 \left( q^2 - \frac{M^2}{\omega} \right)
P_s^{\mn \rho\sg}
\biggr] h_{\rho\sg} \, .
\eea
The gauge fixing action eq. (\ref{eq:gravity_GF}), with flat spacetime
as the background can also be written in the language of 
projectors as follows,
\bea
\label{eq:SgfGR_proj}
&&
S_{\rm GF} = \frac{1}{32 \pi G \al} \int \frac{{\rm d}^dq}{\pd}
h_{\mn} \, q^4 \, \biggl[
-\frac{1}{2} P_1^{\mn \rho\sg} + \frac{1-\bt}{d^2} \biggl\{ 
(1+\rho)^2 (d-1) P_s^{\mn \rho\sg} 
\notag \\
&&
+(d-1-\rho)^2 P_w^{\mn \rho\sg} 
- \sqrt{d-1} (1+\rho)(d-1-\rho)
\left( P_{sw}^{\mn \rho\sg}  + P_{ws}^{\mn \rho\sg}  \right)
\biggr\}
\biggr]
h_{\rho\sg} \, .
\eea
Writing the gauge fixing action in terms of projectors is advantageous. It
shows that it only give contribution to the longitudinal modes of the field, and
not to the transverse spin-2 projector. This is expected as the spin-2 component 
is gauge invariant under the diffeomorphism transformation of the fluctuation
field $h_\mn$. Also it should be noticed that the components corresponding
to each of the projector is of $q^4$ type. This is due to fact that $Y_\mn$ 
contains derivatives. Had we chosen $Y_\mn$ to be proportional to constant,
then we would have got coefficients corresponding to each of projector in 
gauge fixing action to be of $q^2$ type. 
 
When gauge parameters take arbitrary 
values then this will over fix the gauge redundancies, and there 
will be longitudinal terms in the gravity propagator. Such terms may not 
give any contribution to one-loop diagrams (except for quadratic divergent ones)
but at higher loops, contribution from them is unavoidable. However the
$S$-matrix constructed will be gauge-invariant. Its only in 
Landau gauge ($\rho=-1$, $\bt=0$ and $\al=0$) that we don't have
any longitudinal modes in the propagator (and don't have to worry 
about ghosts) describing physical degrees of freedom correctly. 
The gravity propagator in this gauge is given by,
\beq
\label{eq:GR_prop}
D^{\mn \rho\sg} =  (16 \pi G) \biggl[
- \frac{2 M^2 P_2^{\mn \rho\sg}}{q^2(q^2 - M^2)}
+ \frac{2 M^2}{(d-2) \omega} \frac{P_s^{\mn\rho\sg}}{q^2(q^2 - M^2/\omega)}
\biggr] 
= (\D_G^{-1})^{\mn\rho\sg} \, , 
\eeq
where $\D_G^{\mn\al\bt}$ is the inverse propagator for the 
$h_\mn$ field including the gauge fixing and is 
symmetric in $\mn$ and $\al\bt$. In co-ordinate space 
this can be written as,
\beq
\label{eq:gr_invprop}
\frac{1}{2} \int \, {\rm d}^dx \, h_\mn \, \D_G^{\mn\al\bt} h_{\al\bt} \, .
\eeq
The graviton propagator written in eq. (\ref{eq:GR_prop}) can be 
analyzed further by doing partial fraction decomposition. 
This enumerates the various propagating modes of the theory. 
This is given by,
\beq
\label{eq:GR_prop1}
D^{\mn,\al\bt} = 16 \pi G\cdot \Biggl[
\frac{(2P_2 - (2/(d-2)) P_s)^{\mu\nu, \alpha\beta}}{q^2 + i \, \epsilon}
+ \frac{2}{d-2}\frac{(P_s)^{\mu\nu, \alpha\beta}}{q^2 - M^2/\omega + i \epsilon}
- \frac{2 \, (P_2)^{\mu\nu, \alpha\beta}}{q^2 - M^2+ i \epsilon}
\Biggr] \, .
\eeq
We note that the first 
term is the usual massless graviton with two degrees of freedom,
the second term is a massive scalar with mass $M/\sqrt{\omega}$ and 
has one degree of freedom (we call it `Riccion'), while the third term is a massive 
spin-2 mode of mass $M$ with five degrees of freedom (we call it $M$-mode). This mode 
has negative residue and breaks unitarity at tree level. However
when quantum corrections are taken into account, it is shown that the
mass of this mode runs in such a way so that it is always above the 
running energy scale \cite{NarainA1, NarainA2}. 

Having obtained the propagator for the gravitational field, we move 
on to study the gauge field action and its coupling with the gravitational 
field $h_\mn$. In the next subsection we will build up the formalism 
to compute the quantum gravity contribution to the gauge field action. 

\subsection{Formalism}
\label{formal}

Under the decomposition of gauge field in to background and fluctuation
as $A_{\mu}^a = \bar{A}_\mu^a + {\cal A}_{\mu}^a$, the 
expansion of field strength tensor $F_\mn^a$ is given by,
$F_{\mu\nu}^a =
\bar{F}_{\mu\nu}^a + D_{\mu} \mathcal{A}_{\nu}^a -D_{\nu} \mathcal{A}_{\mu}^a  
+ f^{abc} \mathcal{A}_{\mu}^b \mathcal{A}_{\nu}^c $, 
where $\bar{F}_{\mu\nu}^a = \partial_{\mu} \bar{A}_{\nu}^a - \partial_{\nu} \bar{A}_{\mu}^a
+ f^{abc} \bar{A}_{\mu}^b \bar{A}_{\nu}^c$ and 
$D_{\mu} \mathcal{A}_{\nu} = \partial_{\mu} \mathcal{A}_{\nu}^a
+ f^{abc} \bar{A}_{\mu}^b \mathcal{A}_{\nu}^c$.

To compute the one-loop quantum gravity contribution to the running 
of gauge coupling, it is sufficient to expand the gauge field action 
up to quadratic in $h_\mn$ and $\mathcal{A}^a_\mu$ around the 
flat metric. This will give rise to various vertices that will be relevant 
for the one-loop computation and the gauge field propagator. This 
second variation is given by, 
\bea
\label{eq:2ndvargauge}
&& \delta^2 S_{\rm gauge} = - \int \frac{{\rm d}^4x \sqrt{-g}}{8 e^2}
\Biggl[
\biggl\{
\left(\frac{1}{4} h^2 - \frac{1}{2} h_{\mu\nu}h^{\mu\nu} \right)
g^{\mu\alpha}g^{\nu\beta}
- 2 h h^{\mu\alpha} g^{\nu\beta} 
+4 h^{\mu\rho}h_{\rho}{}^{\alpha} g^{\nu\beta}
\notag \\
&&
+ 2 h^{\mu\alpha} h^{\nu\beta}
\biggr\} F_{\mu\nu}^a F_{\alpha\beta}^a
+2 h g^{\mu\alpha}g^{\nu\beta} \left( D_{\mu} \mathcal{A}_{\nu}^a
-D_{\nu} \mathcal{A}_{\mu}^a \right) F_{\alpha\beta}^a
- 8 h^{\mu\alpha} g^{\nu\beta} \left( D_{\mu} \mathcal{A}_{\nu}^a
-D_{\nu} \mathcal{A}_{\mu}^a \right) F_{\alpha\beta}^a
\notag \\
&& 
+ 4 g^{\mu\alpha}g^{\nu\beta} f^{abc} \mathcal{A}_{\mu}^b
\mathcal{A}_{\nu}^c F_{\alpha\beta}^a
+ 2 \eta^{\mu\alpha} \eta^{\nu\beta} 
\left(
D_{\mu} \mathcal{A}_{\nu}^a - D_{\nu} \mathcal{A}_{\mu}^a \right)
\left(
D_{\alpha} \mathcal{A}_{\beta}^a - D_{\beta} \mathcal{A}_{\alpha}^a 
\right)
\Biggr] \, .
\eea
The gauge fixing action for the gauge field given in eq. (\ref{eq:sgf_gauge})
is written for arbitrary background metric. This is crucial so as to have 
diffeomorphism invariance for the effective action obtained by integrating out the 
quantum fluctuations. However for investigating issues related to charge 
renormalization (divergent part of the effective action), it is sufficient to consider 
the flat background. This is justified further by writing the arbitrary background metric
$\bg_\mn$ as $\eta_\mn +H_\mn$, and expanding the gauge fixing action in 
powers of $H_\mn$. The leading term of this series is the same as is obtained when 
background is flat, and is the only relevant term required for studying 
divergent contributions which are zeroth order in curvature of background metric.
This gauge fixing action when added to the second variation of the gauge field action, 
gives the gauge fixed propagator and the vertices, which have been 
written in the appendix \ref{gaugeverprop}. 

Having obtained the expansion of the bare action of gravity and gauge field up
to second order in fields (which is all that is needed in the one-loop 
computation), we start by considering the path-integral over the 
fluctuation fields. The zeroth order term being independent of the fluctuation
fields, can be taken out of the path-integral. The linear term can be removed 
by redefining the fluctuation fields, which only give rise to a trivial Jacobian
from the functional measure. The quadratic piece can now be tackled easily by 
clubbing the two fields to form a multiplet $\Phi = \left( h_{\mu\nu}, \mathcal{A}^a_{\alpha} \right)$.
This allows one to express the second variation in a more 
compact notation. The residual path-integral along with the 
source terms has the following form,
\beq
\label{eq:Z_exp}
Z[\textbf{J}] = \exp\biggl[ i \biggl(S_{\rm GR}(\bg) + S_{\rm gauge}(\bar{A}) 
\biggr)\biggl]
\int \, {\cal D} \Phi
\, \exp \biggl[
\frac{i}{2} \int \, {\rm d}^dx \Phi \cdot \textbf{M} \cdot \Phi^T
+ i \Phi \cdot \textbf{J}^T
\biggr] \, ,
\eeq
where $\textbf{J}=\{ t_\mn, U_\mu^a \}$ is the source multiplet
which couples with the fluctuation fields 
$\Phi = \left( h_{\mu\nu}, \mathcal{A}^a_{\alpha} \right)$, 
and
\bea
\label{eq:M_op}
&& 
\frac{1}{2} \int {\rm d}^d x \, \Phi \cdot \textbf{M} \cdot \Phi^T
= \frac{1}{2} \int {\rm d}^dx \,
\left( \begin{array}{c c}
h_{\mu\nu} & \mathcal{A}^a_{\gamma'}
\end{array}
\right) \cdot \textbf{M} \cdot 
\left( \begin{array}{c}
h_{\rho\sigma} \\
\mathcal{A}^c_{\gamma}
\end{array}
\right)
\notag \\
&&
\textbf{M} = 
\biggl[
\begin{array}{c c}
\Delta_G^{\mu\nu\rho\sigma}
- \frac{1}{4 e^2} V^{\mu\nu\rho\sigma}_{\theta\tau\alpha\beta} \,
\bar{F}^{a\theta\tau} \,\bar{F}^{a\alpha\beta}
& 
-\frac{2}{4e^2} V^{c\mu\nu\gamma'\gamma} D_{\gamma'}
\\
\frac{2}{4e^2} V^{a\rho\sigma\gamma\gamma'} D_{\g}
+ \frac{2}{4 e^2} D_{\g} V^{a\rho\sigma\g\g'}
& 
-\frac{1}{4e^2} \left(
\Delta_g + \Delta_1 + \Delta_2 + \Delta_3 
\right)^{ac\gamma'\gamma}
\end{array}
\biggr] \, .
\eea
From the generating functional $Z$ we define the 
one-particle-irreducible (1PI) generating 
functional $\G = W[\textbf{J}] - \int \, {\rm d}^dx 
\textbf{J} \cdot \langle\Phi^T \rangle$, where $W[\textbf{J}] =
-i \ln Z[\textbf{J}]$ and $\langle \Phi^T \rangle$ is the 
expectation value of $\Phi^T$ field. 
The 1PI generating functional $\G$ is also the effective action. 
Performing the $\Phi$ integration one obtains,
\beq
\label{eq:afterPhi_int}
\exp \biggl[
i \biggl(
\G + \int \, {\rm d}^dx \textbf{J} \cdot \langle \Phi^T \rangle
\biggr)
\biggr]
= \exp \biggl[
i \biggl(
S_{\rm GR}(\bg) + S_{\rm gauge}(\bar{A})
- \frac{1}{4} \int \, {\rm d}^dx 
\textbf{J} \cdot \textbf{M}^{-1} \cdot \textbf{J}^T
\biggr)
\biggr]
\cdot \biggl( \det \, \textbf{M} \biggr)^{-1/2}
\eeq
For zero source, the above equation gives the expression for 
1-loop effective action $\G$,
\beq
\label{eq:EA_1loop}
\G^{\rm 1-loop} [\Phi] = S_{\rm GR} (\bg) + S_{\rm gauge}(\bar{A}) 
+ \frac{i}{2} {\rm Tr} \ln \textbf{M} \, ,
\eeq
where the first two terms correspond to tree level 
diagrams, while the last term contains 1-loop 
quantum corrections. Our task in the following 
is to compute the divergent terms present in 
${\rm Tr} \ln \textbf{M}$ which are proportional 
to $\int \, {\rm d}^dx F_\mn^a F^{a\mn}$. There are 
several ways of obtaining it, but here we will do it by 
expanding the $\ln \textbf{M}$ in a power series, and studying terms
which will contain divergent contribution proportional to $\int \, {\rm d}^dx F_\mn^a F^{a\mn}$.
In our case writing $\textbf{M}= \D + \textbf{V}$ (where the first term contains 
only the propagator of the theory and the second terms 
contains the interactions), we have, 
\beq
\label{eq:M_again}
\textbf{M} = \biggl[
\ba{c c}
\D_G^{\mn\rho\sg} & 0 \\
0 & -\frac{1}{4e^2} \D_g^{ac\g'\g}
\ea 
\biggr]
-\frac{1}{4e^2} \biggl[
\ba{c c}
V^{\mn\rho\sg}_{\ta\tau\al\bt} \bar{F}^{a\ta\tau} \bar{F}^{a\al\bt}
& 
2 V^{c\mn\g'\g}D_{\g'} \\
- 2 V^{a\rho\sg\g\g'}D_{\g} - 2 D_\g V^{\rho\sg\g\g'}
&
\left(\D_1 + \D_2 +\D_3 \right)^{ac\g'\g}
\ea
\biggr] \, ,
\eeq
$\D$ can be pulled outside so that the logarithm of the residual expression 
$\left(\textbf{I} + \D^{-1} \cdot \textbf{V} \right)$ (where $\textbf{I}$ 
is the identity in field space), can be easily expanded as power 
series in $ \D^{-1} \cdot \textbf{V}$. 
\beq
\label{eq:log_exp}
{\rm Tr} \ln \textbf{M} 
= {\rm Tr} \ln \D \cdot \biggl( \textbf{I}
+ \D^{-1} \cdot \textbf{V} \biggr)
= {\rm Tr} \ln \D
+ {\rm Tr} \biggl[
\D^{-1} \cdot \textbf{V}
- \frac{1}{2} \D^{-1} \cdot \textbf{V} \cdot \D^{-1} \cdot \textbf{V}
+ \cdots
\biggr] \, .
\eeq
After pulling out $\D$, ${\bf M}$ has the following 
expression in index form,
\bea
\label{eq:M_alter2}
&& \textbf{M} 
=\biggl(
\ba{c  c}
\D_G^{\mu'\nu',\mn} & 0 \\
0 & -\frac{1}{4e^2} \D_g^{a'a\g'\kp}
\ea
\biggr) \cdot 
\Biggl[
\biggl(
\ba{c c}
\bf{I}^{\rho\sg}_{\mn} & 0 \\
0 & \de^c_{a} \de^\g_{\kp}
\ea
\biggr)
\notag \\
&&
-\frac{1}{4e^2} \biggl(
\ba{c c}
\left(\D_G^{-1}\right)_{\mn}{}^{\al'\bt'}V_{\al'\bt'}{}^{\rho\sg}{}_{\ta\tau\al\bt}
\bar{F}^{b\ta\tau} \bar{F}^{b\al\bt}
&
2 \left(\D_G^{-1}\right)_{\mn}{}^{\al'\bt'} V^c{}_{\al'\bt'}{}^{\g'\g}D_{\g'}
\\
8e^2 \left(\D_g^{-1}\right)_{a\kp}^{b\tau}
\left(V_{b}{}^{\rho\sg\g}{}_{\tau} D_{\g} 
+D_{\g} V_{b}{}^{\rho\sg\g}{}_{\tau} \right) 
&
-4e^2 \left( \D_g^{-1} \right)_{a\kp}^{b\tau}
\left(\D_1 + \D_2 + \D_3 \right)_{b}{}^{c}{}_{\tau}{}^{\g}
\ea
\biggr)
\Biggr]
\eea
In the present case we have taken the background metric to be 
flat, therefore one can ignore the contribution coming from 
first term in the expansion in eq. (\ref{eq:log_exp}), which gives
just a normalization constant. But had we had an arbitrary background
which has a nonzero curvature, then this term cannot be 
ignored and actually give contribution which goes in renormalizing 
the gravitational parameters.

In the following we will compute the one-loop quantum gravity 
corrections to the running of gauge coupling, by evaluating the various
terms of the series in eq. (\ref{eq:log_exp}). 

\subsection{Graphs}
\label{graphs}

We note that the series given in eq. (\ref{eq:log_exp}) 
contains various 1-loop diagrams. Being interested 
in computing the divergent term proportional to $\bar{F}^2$,
the first two terms of the series are sufficient for this. The first term
of the series will give a tadpole type of diagram, with two external
$\bar{F}$ lines and a graviton loop, while the 
second term of series will give a bubble sort of diagram,
which has two external $\bar{F}$ lines, with the loop 
containing one graviton propagator and one gluon 
propagator. In the following we will show how
these two diagrams arise from the series and will compute
the divergent part of both diagrams.

\subsubsection{Tadpole}
\label{tad}

The contribution to the tadpole graph comes from the 
first term of the series in eq. (\ref{eq:log_exp}) which in our case
is $\D^{-1}\cdot \textbf{V}$, where the trace not only acts 
on the field space but also on all the Lorentz and gauge indices.
The contribution of tadpole is given by,
\beq
\label{eq:tad_not}
\G_{{\rm Tad}} = -\frac{i}{8e^2} 
\, {\rm tr} \biggl[
\left(\D_G^{-1}\right)_{\mn}{}^{\al'\bt'}V_{\al'\bt'}{}^{\rho\sg}{}_{\ta\tau\al\bt}
\bar{F}^{b\ta\tau} \bar{F}^{b\al\bt}
-4e^2 \left( \D_g^{-1} \right)_{a\kp}^{b\tau}
\left(\D_1 + \D_2 + \D_3 \right)_{b}{}^{c}{}_{\tau}{}^{\g}
\biggr] \, ,
\eeq
where now the trace `${\rm tr}$' represent the trace over 
un-contracted Lorentz and gauge indices and configuration 
space integration. 
From the expression in eq. (\ref{eq:tad_not}) we note that there 
will be two kind of tadpole diagrams. One in which there is 
graviton in the loop and other in which there is gluon in the loop.
The first set of diagram will give rise to quantum gravity correction 
to the running gauge coupling while the second set of diagrams 
are the usual ones encountered in non-abelian gauge theories 
without gravity, which arise due to self interactions between gluons. 
The diagram giving quantum gravity contribution to the running
gauge coupling is shown in Fig. \ref{fig:1loop}c.
The diagram is computed with $\bar{F}$ as an external leg.
The trace can be written in the co-ordinate space as follows,
\beq
\label{eq:tad_trace}
\G_{{\rm Tad}} =-\frac{i}{8e^2}
\int \, {\rm d}^dx \, {\rm d}^dy \, \left(\D_G^{-1} \right)_{\mn\al'\bt'}(x-y)
\, V^{\al'\bt'\rho\sg}_{\ta\tau\al\bt} \bar{F}^{b\ta\tau}(x)
\bar{F}^{b\al\bt}(y) \, \de(y-x) \de^{\mu}_{\rho} \de^{\nu}_{\sg} \, .
\eeq
Being in flat background this can be re-written in momentum space after 
plugging the expression for the $h$-propagator given in eq. (\ref{eq:GR_prop}).
\bea
\label{eq:tad_mom_exp}
\G_{{\rm Tad}} =
&&
-\frac{i(16 \pi G)}{8e^2} 
\int \, \frac{{\rm d}^d k}{\pd} \, \bar{F}^{b\ta\tau}(k)
\bar{F}^{b\al\bt}(-k) V^{\rho\sg\mn}_{\ta\tau\al\bt} \,
\notag \\
&&
\times \int \, \frac{{\rm d}^dp}{\pd} \biggl[
- \frac{2 M^2 (P_2)_{\mn \rho\sg}}{p^2(p^2 - M^2)}
+ \frac{2 M^2}{(d-2) \omega} \frac{(P_s)_{\mn\rho\sg}}{p^2(p^2 - M^2/\omega)}
\biggr] \, .
\eea
We note that the momentum integration over the integrand is a
Lorentz covariant quantity. This knowledge leads to lot of simplification
specially if we are interested in only the divergent part of the diagrams,
in the sense that under the momentum integration we can replace the 
various spin projectors (which are integration variable dependent) 
by different combinations of $\eta_\mn$ (background 
metric). This is given in more detail in appendix \ref{momint}. Using 
the results of the appendix \ref{momint} 
we obtain the following expression for the tadpole graph,
\bea
\label{eq:tadLorInt}
\G_{{\rm Tad}} =
&&
-\frac{i(\pi M^2 G)}{e^2} 
\int \, {\rm d}^d x \, \bar{F}^{b\ta\tau}\bar{F}^{b\al\bt}
\int \, \frac{{\rm d}^dp}{\pd} \biggl[
\frac{(d-2)(d+1)(d^2-9d+12)}{d(d-1)} \frac{1}{p^2(p^2-M^2)}
\notag \\
&&
+ \frac{(d-5)(d^2-7d+8)}{d(d-1)(d-2)} \frac{\omega^{-1}}{p^2(p^2 - \frac{M^2}{\omega})}
\biggr] \, .
\eea
%

\subsubsection{Bubble Graph}
\label{bub}

The next quantum gravity divergent contribution to gauge coupling beta function
comes from the second term of the series in eq. (\ref{eq:log_exp}). 
For this one needs to compute the 
square of the matrix written in second line of eq. (\ref{eq:M_alter2}). 
Denoting this square of matrix by $N$, its various entries 
are given by,
\bea
\label{eq:F4}
&&
N_{11} = \frac{1}{16e^4} \Biggl[
\left(\D_G^{-1}\right)_{\mu'\nu'\al'\bt'} V^{\al'\bt'\rho'\sg'}
_{\ta\tau\al\bt} \bar{F}^{b\ta\tau}\bar{F}^{b\al\bt}
\cdot \left(\D_G^{-1}\right)_{\rho'\sg'\al''\bt''}V^{\al''\bt''\rho\sg}
_{\ta'\tau'\ta''\tau''}\bar{F}^{d\ta'\tau'}\bar{F}^{d\ta''\tau''}
\Biggr] \, ,
\\
\label{eq:bub1}
&&
N_{12} = \frac{1}{e^2} \Biggl[
\left(\D_G^{-1}\right)_{\mu\nu\al\bt} 
V^{c\al\bt\g'\g} \, D_{\g'}  \left(\D_g^{-1}\right)^{cb}_{\tau\g} 
\biggl( 
V^{b\rho\sg\sg'\tau} D_{\sg'} + D_{\sg'} V^{b\rho\sg\sg'\tau}
\biggr)
\Biggr] \, ,
\\
\label{eq:bub2}
&&
N_{21}=\frac{1}{e^2} \Biggl[
\left(\D_g^{-1}\right)^{ab}_{\tau\kp}
\biggl(
V^{b\rho\sg\g\tau}D_{\g} + D_{\g} V^{b\rho\sg\g\tau}
\biggr) \left(\D_G^{-1}\right)_{\rho\sg\al\bt}
\cdot V^{a\al\bt\g'\ep}D_{\g'}
\Biggr] \, ,
\\
\label{eq:bub_SM}
&&
N_{22} = \Biggl[
\left(\D_g^{-1}\right)_{ab}^{\tau\kp} \left(
\D_1 + \D_2 + \D_J\right)_{\tau\g}^{bc}
\left(\D_g^{-1}\right)_{cb'}^{\al\g} \left(
\D_1 + \D_2 + \D_J\right)^{b'r}_{\al\ep}
\Biggr] \, .
\eea
From this we note that under the trace, $N_{12}$ and $N_{21}$
are same. The contribution of $N_{11}$ is proportional 
to $\bar{F}^4$. In our case of higher derivative gravity this 
gives a finite contribution. 
If we were studying Einstein-Hilbert gravity coupled with gauge field,
then this term would be $log$-divergent and will give rise to a 
counter term proportional to $F^4$, signaling the non-renormalizabilty 
of the coupled gauge-gravity Lagrangian. 
The contribution coming from $N_{22}$ does not contains 
any quantum gravitational correction, as there are no graviton loops.
It gives rise to the same diagrams and terms that one witnesses 
in non-abelian theories without gravity. Only $N_{12}$ and
$N_{21}$ contain divergent quantum gravitational contributions to the $\bar{F}^2$ terms.
These are bubble kind of diagrams with the loop containing 
one graviton propagator and one gluon propagator, as shown in Fig. \ref{fig:1loop}d. 
The contribution of the bubble diagram is given by,
\beq
\label{eq:bub_not}
\G_{{\rm Bub}}= -\frac{i}{2e^2}
{\rm tr} \biggl[
\left(\D_G^{-1}\right)_{\mu\nu\al\bt} 
V^{c\al\bt\g'\g} \, D_{\g'}  \left(\D_g^{-1}\right)^{cb}_{\tau\g} 
\biggl( 
V^{b\rho\sg\sg'\tau} D_{\sg'} + D_{\sg'} V^{b\rho\sg\sg'\tau}
\biggr)
\biggr]\, ,
\eeq
where $V^{c\al\bt\g'\g}$ and $V^{b\rho\sg\sg'\tau}$ are vertices 
given in appendix \ref{gaugeverprop}.
The trace can be expanded and written in co-ordinate space as,
\bea
\label{eq:bub_cod}
&&
\G_{\rm Bub} = -\frac{i}{2e^2} 
\int \, {\rm d}^dx \, {\rm d}^dy \, {\rm d}^dz \,
\left(\D_G^{-1}\right)_{\mn\al\bt}(x-y) \cdot 
V^{c\al\bt\g'\g}(y) \, D^{y}_{\g'} \cdot \left(\D_g^{-1}\right)^{cb}_{\tau\g}(y-z) 
\notag \\
&& \times \biggl(
V^{b\rho\sg\sg'\tau}(z) D^{z}_{\sg'} + D^{z}_{\sg'} V^{b\rho\sg\sg'\tau}(z)
\biggr) \, \de(z-x) \, .
\eea
Here the two derivatives $D$ that appear are covariant 
derivatives constructed with the background gauge field. These can be 
expanded in partial derivative piece plus the background gauge field piece. 
It should be noted that the divergent contribution form this diagram
comes only from the piece when both the covariant derivative 
have become partial derivative. This can be singled out easily and 
in momentum space has the following expression.
\beq
\label{eq:bub_mom}
\G_{{\rm Bub}} = \frac{i}{2e^2} 
\int {\rm d}^dx V^{c\al\bt\g'\g}V^{b\rho\sg\ta\tau}
\int \frac{{\rm d}^dp}{\pd} \left(\D_G^{-1}\right)(p)_{\rho\sg\al\bt}
\left(\D_g^{-1}\right)(p-k)^{cb}_{\tau\g} 
(p-k)_{\g'}  (p-k)_{\ta} \, ,
\eeq
where $\left(\D_g^{-1}\right)(p-k)^{cb}_{\tau\g} $
is the gauge propagator in momentum space 
carrying the momenta $(p-k)$. Its expression is the 
following,
\beq
\label{eq:gluon_prop_mom}
\left(\D_g^{-1}\right)(p-k)^{cb}_{\tau\g} 
= \frac{\de^{cb}}{4(p-k)^2} \biggl[
\eta_{\tau\g} - (\xi -1) \frac{(p-k)_{\tau}(p-k)_{\g}}
{(p-k)^2}
\biggr] \, .
\eeq
The vertices $V^{c\al\bt\g'\g}$ and $V^{b\rho\sg\ta\tau}$
are symmetric in $\al\bt$ and $\rho\sg$ respectively, while 
they are anti-symmetric in $\g'\g$ and $\ta\tau$ respectively. On plugging the
gauge propagator from eq. (\ref{eq:gluon_prop_mom}) 
in the bubble contribution, it is noted that the term proportional 
to $(\xi-1)$ being symmetric in pairs $\g'\g$ and
$\tau\ta$ respectively cancel due anti-symmetry property 
of the vertex $V$ in the last two index (check appendix \ref{gaugeverprop}). 
This clearly shows that the contribution from this diagram will be independent of the 
gauge fixing condition for the gauge field. This implies the 
expression,
\beq
\label{eq:bub_simp}
\G_{{\rm Bub}} = \frac{i}{8e^2}
\int {\rm d}^dx V^{c\al\bt\g'\g} V^{b\rho\sg\ta\tau}
\int \frac{{\rm d}^dp}{\pd} \left(\D_G^{-1}\right)(p)_{\rho\sg\al\bt}
\frac{(p-k)_{\g'}(p-k)_{\ta}}{(p-k)^2} \eta_{\tau\g} \, .
\eeq
In order to isolate just the divergent part of this, we consider 
only the momentum integral. This momentum integral is a completely 
Lorentz covariant quantity constructed using the background metric 
$\eta_\mn$ and the external momenta $k_\mu$. Denoting this by $I(k)$ (with 
appropriate indices), we have the following expression for it.
\beq
\label{eq:IK}
I(k)_{\rho\sg\al\bt\g'\ta} = \int \frac{{\rm d}^dp}{\pd}
\left(\D_G^{-1}\right)(p)_{\rho\sg\al\bt}
\frac{(p-k)_{\g'}(p-k)_{\ta}}{(p-k)^2} \, .
\eeq
$I(k)$ being analytic in $k_\mu$, can be expanded in powers of $k_\mu$
around $k_\mu=0$, where each coefficient in series is obtained by 
taking successive derivatives of $I(k)$ with respect to $k_\mu$ and setting $k_\mu=0$.
Depending on the gravitational propagator, only a finite number of coefficients will have 
divergence. It is to be noted that the leading term of this 
expansion $I(0)$ will be proportional to $\bar{F}^2$.
In our case of higher derivative gravity, only this term contains 
the divergence and is $log$-divergent (in case of Einstein-Hilbert
gravity, the next two terms of series will also have divergences). Plugging the higher derivative 
gravity propagator in the expression of $I(0)$ we obtain,
\bea
\label{eq:I0}
I(0)_{\rho\sg\al\bt\g'\ta} 
&=& \int \frac{{\rm d}^dp}{\pd}
\left(\D_G^{-1}\right)(p)_{\rho\sg\al\bt}
\frac{p_{\g'}p_{\ta}}{p^2}
\notag \\
&=& 16 \pi G \int \frac{{\rm d}^dp}{\pd} 
\biggl[
- \frac{2 M^2 (P_2)_{\rho\sg\al\bt}}{p^2(p^2 - M^2)}
+ \frac{2 M^2}{(d-2) \omega} \frac{(P_s)_{\rho\sg\al\bt}}{p^2(p^2 - M^2/\omega)}
\biggr] 
\times \frac{p_{\g'}p_{\ta}}{p^2} \, .
\eea
Under the momentum integration the tensor structure present in the 
integrand can be replaced by most general tensor that can 
be constructed using flat spacetime metric $\eta_\mn$ obeying all 
the symmetries of the integrand. This tensor structure constructed with 
$\eta_\mn$ has been worked out in the appendix \ref{momint}. Plugging this 
in the contribution of the bubble diagram we get the following,
\bea
\label{eq:bub_hdg_momnt}
\G_{{\rm Bub}} = 
&&
\frac{i \, 4 \pi M^2 G}{e^2} \int \frac{{\rm d}^dp}{\pd}
\biggl[
-\frac{2(d-2)(d+1)}{d(d-1)} \frac{1}{p^2(p^2-M^2)}
\notag \\
&&
+ \frac{1}{\omega} \frac{(d-3)^2}{d(d-1)}
\frac{1}{p^2(p^2- \frac{M^2}{\omega})}
\biggr]
\int {\rm d}^dx \, {\rm tr} F^2 \, .
\eea
%

\subsubsection{$4-\ep$ Regularization}
\label{4Dreg}

Having obtained the expression of the tadpole and bubble graph in arbitrary 
dimensions (without the momentum integration), we perform the 
momentum integration. Choosing a regularization scheme which doesn't
interfere with the gauge invariance of the theory in important. 
Dimensional regularization is an ideal choice which isolates the divergent piece 
very clearly. On performing the momentum integration in arbitrary dimensions,
the contribution of tadpole and bubble diagram in arbitrary dimensions is the following,
\bea
\label{eq:tadbub_d}
\G_{{\rm Tad}} 
&=& 
-\frac{2 \pi M^2 G}{e^2 (4 \pi)^{\frac{d}{2}}}
\G\left(2 - \frac{d}{2} \right) \biggl[
\frac{(d+1)(d^2-9d+12)}{d(d-1)} M^{d-4}
\notag \\
&&
+ \frac{1}{\omega} \frac{(d-5)(d^2-7d+8)}{d(d-1)(d-2)^2} 
\left(\frac{M^2}{\omega}\right)^{\frac{d}{2}-2}
\biggr] \int {\rm d}^dx {\rm tr} F^2 \, ,
\notag \\
\G_{{\rm Bub}} 
&=& 
\frac{8 \pi M^2 G}{e^2 (4\pi)^{d/2}}
\G\left(2 - \frac{d}{2} \right) \biggl[
-\frac{2(d+1)}{d(d-1)} M^{d-4} 
\notag \\
&&
+ \frac{1}{\omega} \frac{(d-3)^2}{d(d-1)(d-2)^2} 
\left( \frac{M^2}{\omega} \right)^{\frac{d}{2}-2}
\biggr] \int {\rm d}^dx \, {\rm tr} F^2 \, .
\eea
The $1/\ep$ pole of these diagrams in the $4-\ep$ dimensional 
regularization scheme is,
\beq
\label{eq:tadbubd4}
\G^{\rm Div}_{\rm Tad} = -\G^{\rm Div}_{\rm Bub} =
\frac{1}{\ep} \frac{M^2 G}{48 \pi \, e^2} \left(40 - \frac{1}{\omega} \right) 
\int {\rm d}^4 x \, {\rm tr} F^2 \, .
\eeq
From this we clearly see that the $1/\ep$ term from both 
the diagrams cancel each other. Thus there is no quantum 
gravity correction at one-loop to the running of gauge coupling.
This means that to one-loop $a=0$ for higher derivative gravity. 
But there is a finite quantum gravity contribution which is 
generated in four dimensions. This is given by,
\beq
\label{eq:EAfiniteD4}
\G_{\rm Grav}^{\rm d=4} = -\frac{M^2 G}{e^2} \frac{13 + 20 \omega}
{192 \pi \omega} \int {\rm d}^4 x \, {\rm tr} F^2 \, .
\eeq
This can be relevant when the theory has no divergent contributions
at all. For example abelian gauge theories without matter fields.
But in realistic situation even the presence of electrons give 
divergent contributions in four dimensions resulting in the running 
of gauge coupling. Still, it is interesting to 
note the nature of this finite renormalization which happens for the 
abelian gauge coupling in the absence of matter fields.
This finite renormalization is given by,
\beq
\label{eq:finiteD4_RG}
\frac{1}{e^2_R} = \frac{1}{e^2} \left[
1+ \frac{M^2 G}{48 \pi} \left( 20 + \frac{13}{\omega} \right)
\right] \, ,
\eeq
which tells that due to quantum gravity correction, the gauge coupling
decreases. 

Coming back to the cancellation of divergent piece, we note that  this 
cancellation of divergent parts of the two diagrams in the case of 
higher-derivative gravity was also observed in \cite{Fradkin, Fradkin1, NAgauge}. Such 
cancellation was also found in the case of EH gravity \cite{Deser1,
Deser4, Deser5, Deser6, Pietrykowski, Toms:2007, Rodigast2008, Folkerts} 
in different regularization schemes
implying the vanishing of `$a$'-term in the gauge beta function.
However in these cases of effective field theory\cite{Donoghue1, 
Donoghue2, Donoghue3}, no clear meaning 
should be associated to these terms \cite{Donoghue4}.
But contemplating on these observations, it is natural to 
ask whether this is an accidental cancellation or if there 
is some deeper symmetry at work responsible for vanishing of
`$a$'-term. This issue is all the more strongly voiced in a 
renormalizable, unitary quantum gravity theory such as the one given by
eq. (\ref{eq:hdg_act}) \cite{NarainA1, NarainA2} (where criticism (\cite{Donoghue4})
as in effective theories \cite{Donoghue1, Donoghue2, Donoghue3} do not arise).
In the section \ref{duality} we study this particular issue in more 
detail. In the next section we will study the
gauge field contribution to the gravitational beta functions 
and will analyze the unitarity issues in this coupled system.
\footnote{
In three dimensions, the coupled theory is finite
as there are no ultraviolet divergences. 
However, there is finite renormalization of couplings.
For electromagnetic coupling, quantum gravity effects 
tend to decrease the strength of coupling.
}

\section{Gravitational Beta functions}
\label{gravbeta}

In this section we will study the running of the gravitational couplings, 
whose beta functions gets corrected due to the presence of 
gauge fields in the loop. These have been computed and studied 
at several places in the literature \cite{Fradkin, Fradkin1, Buchbinder, 
Avramidibook, Gorbar, Shapiro},
here we will analyze them to study the unitarity of the coupled gauge-gravity 
system in light of the recent work on the quantum theory of higher-derivative 
gravity \cite{NarainA1, NarainA2}. The running of couplings are studied 
around the gaussian fixed point and the unitarity of the flow is analyzed 
within the perturbation theory. 

Elsewhere higher-derivative gravity has been studied from a different perspective
 \cite{Codello2006, Benedetti2009, Niedermaier2009}.
There renormalization group flow was obtained using non-perturbative methods \cite{Wetterich},
and the unitarity of the flow was investigated around a non-trivial fixed point,
within the realm of asymptotic safety scenario 
\cite{Weinberg, AS_rev1, AS_rev2, AS_rev3, AS_rev4}. In this paper
we follow the perturbative methods and study the flow
around the gaussian fixed point.

The beta function of the gravitational couplings in $4-\ep$ dimensional 
regularization scheme in the Landau gauge are following
\cite{Buchbinder, Avramidibook, Gorbar},
\bea
&&
\label{eq:M2Gbeta}
\frac{{\rm d}}{{\rm d} t} \left(\frac{1}{M^2 G} \right)
= -\frac{1}{\pi} \left[\frac{133}{10} + \frac{1}{5}N_A \right] \, ,\\
&&
\label{eq:wM2Gbeta}
\frac{{\rm d}}{{\rm d} t} \left( \frac{\omega}{M^2 G} \right)
= \frac{1}{\pi} \left[\frac{5}{3} \omega^2 + 5 \omega + \frac{5}{6} \right] \, ,\\
&&
\label{eq:Gbeta}
\frac{{\rm d}}{{\rm d} t} \left(\frac{1}{G} \right)
= \frac{M^2}{\pi} \left(\frac{5}{3} \omega - \frac{7}{24 \omega} \right) \, ,
\eea
where $N_A$ is the number of gauge boson in the theory
\cite{Fradkin, Fradkin1, Buchbinder, Avramidibook, Gorbar, Shapiro}. 
For abelian gauge theory $N_A=1$, for non-abelian gauge theory like $SU(N)$, $N_A=N^2-1$. 
These beta functions now can be analyzed along the same lines as in 
\cite{NarainA1, NarainA2}. Using eq. (\ref{eq:M2Gbeta} and \ref{eq:wM2Gbeta}) one 
can extract the flow of the parameter $\omega$. This is given by,
\beq
\label{eq:wbeta}
\frac{{\rm d} \omega}{{\rm d} t} =
\frac{M^2G}{\pi} \left[
\frac{5}{3} \omega^2 + \left(\frac{183}{10} + \frac{1}{5} N_A \right) \omega
+ \frac{5}{6}
\right] 
= \frac{5 M^2 G}{3 \pi}  (\omega + \omega_1)(\omega + \omega_2)\, ,
\eeq
where $\omega_1$ and $\omega_2$ are given by following,
\bea
\label{eq:wroots}
&&
\omega_1=\frac{1}{100} [(549+6N_A) - \sqrt{(549+6N_A)^2 -5000}] \, ,
\notag \\
&&
\omega_2=\frac{1}{100} [(549+6N_A) + \sqrt{(549+6N_A)^2 -5000}] \, .
\eea
For all values of $N_A$ both $\omega_1$ and $\omega_2$ are positive. From the
rhs of eq. (\ref{eq:wbeta}) we note that the beta function has two fixed points
$-\omega_1$ (UV repulsive) and $-\omega_2$ (UV attractive). From the gravity 
propagator in eq. (\ref{eq:GR_prop1}) it is clearly noted that only positive values
of $\omega$ are allowed and considered physical, we therefore realize that 
both these fixed points lie in the unphysical domain.
For all positive values of $\omega$, the rhs of eq. (\ref{eq:wbeta}) is always 
positive, thereby implying that $\omega$ is a monotonic increasing 
function of $t$ and vice versa. Eq. (\ref{eq:M2Gbeta}) readily allows us 
to express $M^2G$ in terms of $t$, with which we can integrate the 
equation of $\omega$ to obtain,
\beq
\label{eq:tw}
t= T
\Biggl[
1- 
\left( \frac{\omega + \omega_2}{\omega + \omega_1}
\cdot 
\frac{\omega_0 + \omega_1}{\omega_0 + \omega_2} \right)^{\alpha}
\Biggr] \, .
\eeq
Calling $U=(133/10 + 1/5 N_A)$, we have $T=\pi /(U M_0^2 G_0)$
and $\alpha= 3U /5(\omega_2 - \omega_1)$, with the subscript 
`$0$' meaning that the coupling parameters are evaluated 
at $t=0$ or $\mu=\mu_0$. Due to monotonic relation between 
$t$ and $\omega$, one can transform any evolution in $t$,
into evolution in $\omega$. Using this the flow of $G$ can readily be written 
in $\omega$-space, 
\beq
\label{eq:betalogG}
\frac{{\rm d} \ln G}{{\rm d} \omega} 
= - \frac{\omega - \frac{7}{40 \omega} }
{(\omega + \omega_1) (\omega + \omega_2)} \, .
\eeq
The flow of $G$ has two fixed points, one is at $\omega=\sqrt{7/40}$
and other at $\omega=\infty$. Both these fixed points lie
in the physical domain. At the former $G$ is maximized, while at 
later $G\to0$. The flow of $G$ can be solved easily in $\omega$-space. This
is given by,
\beq
\label{eq:Gflow}
\frac{G}{G_0} = \frac{\omega_0}{\omega} \cdot 
\left(
\frac{1 + \omega_1/\omega }
{1 + \omega_1/\omega_0 }\right)^{A1}
\left(
\frac{1 + \omega_2/\omega }
{1 + \omega_2/\omega_0 }\right)^{A2}  \, ,
\eeq  
where,
\bea
\label{eq:GAs}
&&
A1=\frac{3 \left(2379 + 26N_A - 9 \sqrt{(6 N_A+549)^2-5000}\right)}
{40 \sqrt{(6 N_A+549)^2-5000}}\, ,
\notag \\
&&
A2 = -\frac{3 \left(2379 + 26N_A + 9 \sqrt{(6 N_A+549)^2-5000}\right)}
{40 \sqrt{(6 N_A+549)^2-5000}} \, .
\eea
From this we see that for large $\omega$, $G \sim 1/\omega$ 
thereby going to zero for large $t$, while
for small $\omega$, $G \sim \omega^{7/20}$
reaching a peak at $\omega_0=\sqrt{7/40}$. Similarly
using eq. (\ref{eq:wM2Gbeta} and \ref{eq:Gbeta}) we can extract
the running of Riccion mass $M^2/\omega$, which along with
eq. (\ref{eq:wbeta}) can be integrated to get the flow of Riccion mass
to be,
\beq
\label{eq:M2byw}
\frac{M^2}{\omega} = \frac{M_0^2}{\omega_0} 
\left(
\frac{1 + \omega_1/\omega }
{1 + \omega_1/\omega_0 }\right)^{B1} 
\left(
\frac{1 + \omega_2/\omega }
{1 + \omega_2/\omega_0 }\right)^{B2} \, ,
\eeq
where,
\bea
\label{eq:MBs}
&&
B1 = \frac{3 \left(2941 +54N_A + 9 \sqrt{(6 N_A+549)^2-5000}\right)}{40 \sqrt{(6 N_A+549)^2-5000}} \, ,
\notag \\
&&
B2 = -\frac{3 \left(2941 +54N_A - 9 \sqrt{(6 N_A+549)^2-5000}\right)}{40 \sqrt{(6 N_A+549)^2-5000}} \, .
\eea
From eq. (\ref{eq:M2byw}) we can now analyze the behavior of $M^2/\mu^2$.
Alternatively we can use eq. (\ref{eq:M2Gbeta}, \ref{eq:Gbeta} and 
\ref{eq:wbeta}) to obtain the running of $M^2/\mu^2$ in the 
$\omega$-space. This is given by, 
\beq
\label{eq:M2bymu}
\frac{{\rm d}}{{\rm d} \omega} \ln \left( \frac{M^2}{\mu^2} \right)
= \frac{ \omega^2 - \frac{399+6N_A}{50} \omega -\frac{7}{40} - \frac{6 \pi \omega}{5 M^2 G}}
{\omega (\omega + \omega_1)(\omega + \omega_2)} \, .
\eeq
The rhs of eq. (\ref{eq:M2bymu}) vanishes for a particular value
of $\omega=\omega_*$ given by, 
\beq
\label{eq:mincond}
\omega_*^2 - \frac{399+6N_A}{50} \omega_* -\frac{7}{40} 
= \frac{6 \pi \omega_*}{5 M_*^2 G_*} \, .
\eeq
The positivity of second derivative of $M^2/\mu^2$ with respect to $\omega$ 
at the point $\omega_*$ tells that this is a minima. By demanding that 
$M^2/\mu^2 >1$ at this minima, we make sure that $M^2/\mu^2$ 
remains greater than one throughout the whole physically allowed range 
of $\omega$. Therefore the $M$-mode is not realizable throughout the 
flow. This condition is easily achievable,
by appropriately choosing $\mu^2_* G_*$. Perturbative loop expansion requires that 
$M^2G$ is small. Therefore $M$ is a sub-Planckian mass,
yet the running mass as dictated by quantum theory makes it
physically not realizable even in post Planckian regime. Similarly using 
eq. (\ref{eq:wM2Gbeta}, \ref{eq:Gbeta} and \ref{eq:wbeta}) we obtain
flow of $M^2/(\omega \mu^2)$,
\beq
\label{eq:M2bywmu2}
\frac{{\rm d}}{{\rm d} \omega} \ln \left(\frac{M^2}{\omega \mu^2} \right)
=- \frac{\frac{27}{40} + \frac{208 + 2N_A}{25} + \frac{6 \pi \omega}{5M^2 G}}{\omega (\omega+\omega_1)
(\omega+\omega_2)} \, ,
\eeq
showing that the Riccion mass relative to $\mu$
decreases monotonically, as the \textit{rhs} of
eq. (\ref{eq:M2bywmu2}) is always positive.
By a suitable choice we can make the
Riccion to be physically realizable or not. So we conclude that 
there exists unitary physical subspace only 
with the gravitons or along with Riccions. This was also observed in the 
case of pure gravity without matter \cite{NarainA1, NarainA2}, we note
that the presence of gauge fields don't change the qualitative 
picture, although the location of various fixed points have been changed. 

Allowed physical range of $\omega$ being between zero and 
infinity, puts a lower and upper bound on the value of $t$.
The range of $t$ is given by,
\beq
\label{eq:tbound}
t_{\rm min} \equiv T
\Biggl[
1- 
\left( \frac{\omega_2}{\omega_1}
\cdot 
\frac{\omega_0 + \omega_1}{\omega_0 + \omega_2} \right)^{\alpha}
\Biggr] \leq t
\leq T
\Biggl[
1- 
\left(  
\frac{\omega_0 + \omega_1}{\omega_0 + \omega_2} \right)^{\alpha}
\Biggr] \equiv t_{\rm max} \, .
\eeq
Our one-loop analysis shows that there is no solution for 
$\omega$ from the eq. (\ref{eq:tw}) for $t>t_{\rm max}$. This 
could be an one-loop artifact and would perhaps go away 
when higher-loop corrections are taken into account, which might 
push $t_{\rm max}$ to infinity. For $t<t_{\rm min}$, $\omega$ becomes
negative, thereby implying that the mass of the Riccion $M/\sqrt{\omega}$
becomes imaginary. This signals the instability of the vacuum. It
is an infrared issue and needs to be considered separately, namely
the effects of cosmological constant. 

\begin{figure}
\centerline{
\begin{minipage}[t]{3.5in}
\vspace{0pt}
\centering
\includegraphics[width=3.5in]{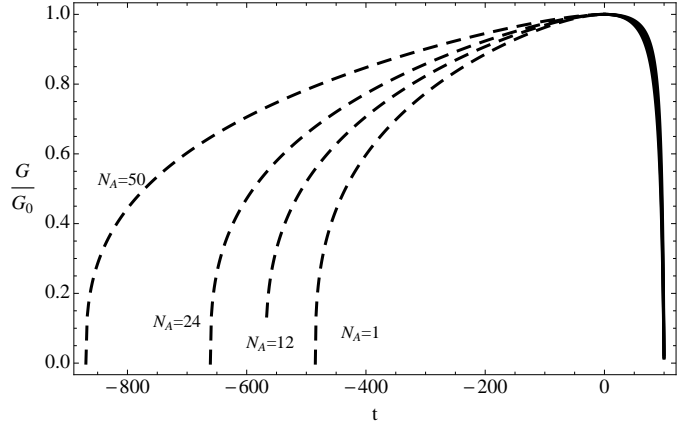}
\end{minipage}
}
 \caption[]{
Running Newton's gravitational coupling for various number of gauge bosons.
Here we are plotting the running of $G$ for $N_A=1$ ($U(1)$ theory),
$N_A=12$ (standard model), $N_A=24$ ($SU(5)$ GUT) and $N_A=50$.
  }
\label{fig:Gt}
\vspace{-5mm}
\end{figure}

To solve the flow of the gravitational parameters, we need some 
initial conditions and the number of gauge boson in the system.
We consider four situations, each with different number of
gauge bosons. We consider case for $U(1)$ gauge theory,
which has $N_A=1$; standard model case with $N_A=12$,
$SU(5)$ GUT which has $N_A=24$ and some other special 
theory which has $N_A=50$. Having fixed 
the $N_A$, we need to fix three initial conditions, as we have the flow equation 
for three parameters. Having already chosen $\omega_0$ 
(the point at which $G$ maximizes) as the reference point,
we need to choose two more parameters in order to completely
fix the renormalization group trajectory of the gravitational sector.
As $M_0^2G_0$ is inversely related to $t_{\rm max}$, thus 
choosing one fixes the other. It is more wise to choose $t_{\rm max}$,
as it tells the number of e-folds that are present between the reference 
point and the point at which $G\to0$. This should be large enough
to incorporate all the known physics.
For the third condition we choose the value of $M/\mu$ at the 
$\omega_*$. This is important to choose appropriately as we don't 
want $M$-mode to be physically realizable throughout the flow. 

For the purpose of quantitative study we choose $t_{\rm max}=100$.
As the flow $G$ is independent of the $M_*/\mu_*$, therefore 
we plot the running of $G$ for various values of $N_A$. It is noted
that for fixed $t_{\rm max}$, increasing $N_A$ decreases $t_{\rm min}$.
The flow of $G$ in the UV for various values of $N_A$ remains same while
the difference is substantial between various RG trajectory of $G$,
in the low energy regime. The flow of $G$ is depicted in Fig. \ref{fig:Gt}.

\begin{figure}
\centerline{
\begin{minipage}[t]{3.5in}
\vspace{0pt}
\centering
\includegraphics[width=3.5in]{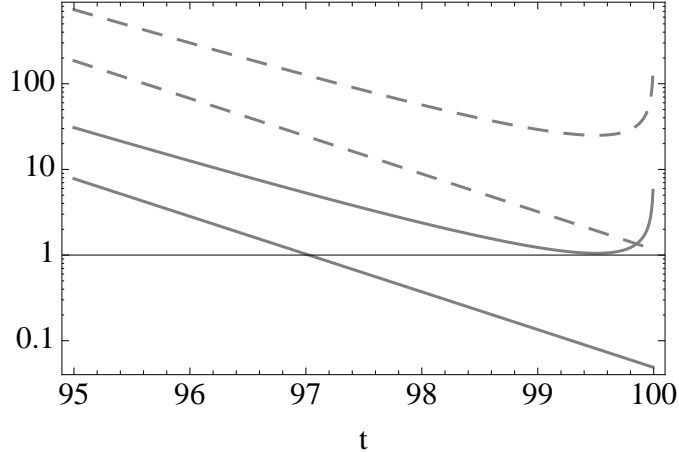}
\end{minipage}
}
 \caption[]{
$M$-mode ratio $M/\mu$ (top) and Riccion mass ratio $M/\mu \sqrt{\omega}$ (bottom).
Dashed lines for $M_*/\mu_*=25$ and solid lines for $M_*/\mu_*=1.05$.
}
\label{fig:ut}
\vspace{-5mm}
\end{figure}

We consider two scenarios one in which Riccion is realizable and
other in which it is not realizable. It turns out there is bound on
$M_*/\mu_*$ for each value of $t_{\rm max}$ and $N_A$, below which Riccion 
is realized in the flow, and above that bound Riccion is also outside the 
spectrum throughout the flow. This bound is given by,
\beq
\label{eq:riccionbound}
\left . \frac{M_*}{\mu_*} \right |_{B}
= e^{t_{\rm max} - t_*} \sqrt{\omega_*}
\left(1+ \frac{\omega_1}{\omega_*} \right)^{B1/2}
\left(1+ \frac{\omega_2}{\omega_*} \right)^{B2/2} \, .
\eeq
This bound depends on $N_A$ and decreases as $N_A$ 
increases, but the dependence is very mild. 
For $t_{\rm max}=100$ and $N_A=1$, this 
bound is given by $M_*/\mu_*=21.51$. Thus we consider two 
cases: $M_*/\mu_*=1.05$ and $M_*/\mu_*=25$. In former 
Riccion is realized while in later Riccion is not realized throughout 
the whole RG trajectory for a given $N_A$. 

For the above values of $M_*/\mu_*$, the flow of the $M^2/\mu^2$
and $M^2/(\omega \mu^2)$ has been plotted in Fig. \ref{fig:ut}.
For each case we note that $M$-mode is never realized throughout the
flow, while Riccion is realized in one case and not in other. For
$t_{\rm max}=100$, choosing $M_*/\mu_*<21.51$, we have Riccion
getting realized. In the Fig. \ref{fig:ut} we have considered the case
when $M_*/\mu_*=1.05$ (solid lines), where we clearly see that
Riccion is getting realized. For $M_*/\mu_*>21.51$, both 
$M$-mode and Riccion is out of physical spectrum throughout 
the whole flow. In Fig. \ref{fig:ut} we considered the case 
when $M_*/\mu_*=25$ (dashed lines), where both $M$-mode and
Riccion is out of physical spectrum. From Fig. \ref{fig:ut} we notice that 
Riccion is physically realizable in high energy scattering processes 
at most for about four e-folds before $t_{max}=100$.

Throughout the whole flow within the physically allowed region 
(between $t_{\rm min}$ and $t_{\rm max}$), 
$M^2G$ remains small and in perturbative regime. 
However the beta function $M^2G$ is such that ultimately its flow 
hits a Landau singularity. Though this occurs way beyond $t_{\rm max}$. The existence of 
Landau singularity could be an artifact of one-loop, and might perhaps go away 
when higher loop corrections are incorporated. This is so because 
at higher loops the beta function of the coupling $1/M^2G$ gets 
a correction which is proportional to couplings instead of a constant,
thereby raising the speculation that there might be a fixed point
for $M^2G$. 

\begin{figure}
\centerline{
\vspace{0pt}
\centering
{\includegraphics[width=3 in]{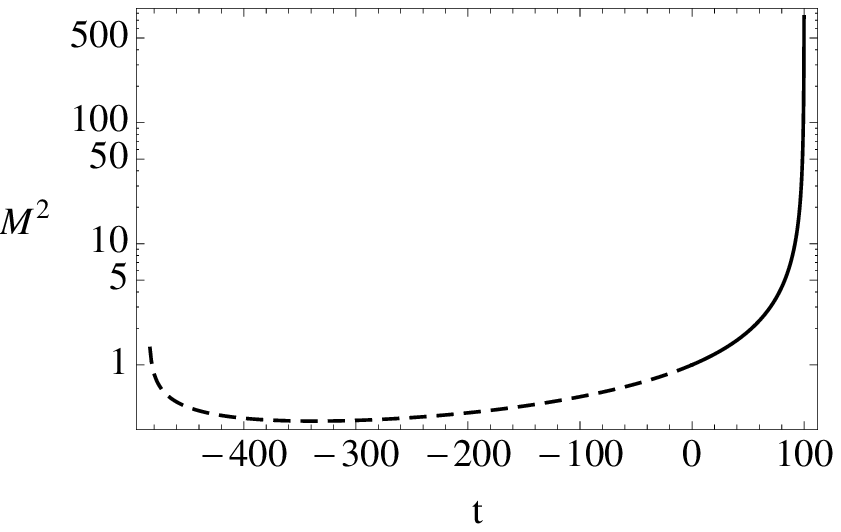}\hspace{10mm}
\includegraphics[width=3.1in]{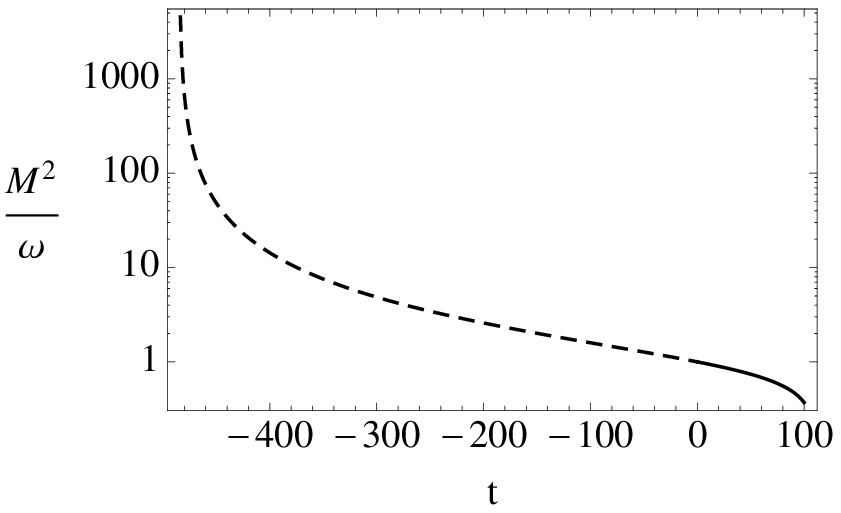}}
}
 \caption[]{
 Flow of $M^2$ and $M^2/\omega$ with respect to $t$, appropriately normalized 
 with respect to the value of $M^2$ and $M^2/\omega$ at reference point $\omega_0$.
}
\label{fig:hdflow}
\vspace{-5mm}
\end{figure}

The flow of $M^2$ and $M^2/\omega$ (the relative coefficients of 
higher-derivative terms with respect to Einstein-Hilbert term) is such 
that in infrared regime both $M^2$ and $M^2/\omega$ goes to infinity, thereby 
suppressing any higher-derivative terms, giving the Einstein-Hilbert action 
in the low energy regime. This can be seen more clearly from Fig. \ref{fig:hdflow}, 
where the flows of $M^2$ and $M^2/\omega$ has been depicted. 
This is important and crucial as then all the known low energy 
physics is reproduced. In the UV, $M^2$ goes to infinity when 
$t \to t_{\rm max}$, signaling that the $M$-mode goes out of
physical spectrum. However the Riccion mass $M^2/\omega$
goes to zero, implying that the contributions from the $R^2$ 
term in the effective action becomes more prominent in the UV.

Having studied the flow of the gravitational parameters in the coupled system 
of gauge field with higher-derivative gravity, we set back to analyze the 
quantum gravity correction to the gauge beta function in more detail. In general
in a coupled system of gravity with gauge fields, the gauge beta function necessarily 
has the structure written in eq. (\ref{eq:beta_gauge_gr}), which automatically arises 
in perturbative study of Feynman path-integral of the coupled system. It is found that
to one-loop $a=0$ in higher-derivative gravity and is independent of the gauge group.
This observation has also been made in a similar one-loop computation in the 
context of Einstein-Hilbert gravity. It is indeed compelling to ask whether there is 
some deeper symmetry principle at work, responsible for vanishing of `$a$'-term? 
As this term is universal to all gauge theories (meaning independent of the gauge coupling),
and is the most dominant term for small coupling, thus it is important 
to understand the nature of this term in more detail by isolating it in the 
context of abelian gauge theories without matter, where `$b$' term 
of the beta function is absent to all loops. Simplicity 
of this coupled system will allow us to gain more insight in to the cause of the 
cancellation of divergences, leading to the vanishing of `$a$'-term in the gauge 
beta function. Therefore in the next section we 
will study the `$a$' term in abelian gauge theory to all loops.

\section{Duality transformation}
\label{duality}

In this section we study abelian gauge theory coupled with gravity in order to 
probe into the cause of vanishing of `$a$' term in the gauge beta function.
Here we give a formal argument to show that $a=0$ due to the self-duality 
property of the abelian gauge theories in four dimensions \cite{Savit, Peskin, NAgauge}.
In this proof gravity action plays no significant
role, so here we don't explicitly specify the gravitational action.
However it is important to mention that the path-integral considered 
should be renormalizable. We start by considering the
path-integral of gauge fields coupled with gravity,
\beq
\label{eq:pathint_em}
Z = \int {\cal D} g_\mn \, {\cal D} A_\rho 
e^{i S_{\rm GR}} \, e^{i S_{\rm EM}} \, ,
\eeq
where $S_{\rm EM} = \left(-1/4e^2\right) \int {\rm d}^4 x \, \sqrt{-g}
\, g^{\mu\al} g^{\nu\bt} F_\mn F_{\al\bt}$, and 
$F_\mn= \partial_\mu A_\nu - \partial_\nu A_\mu$ and $S_{\rm GR}$
is some arbitrary renormalizable and unitary gravity action. This is important 
as otherwise the path-integral will not be well-defined perturbatively.
For this theory the running of the gauge coupling $e^2$ is given by the
following beta function,
\beq
\label{eq:e2beta}
\frac{{\rm d}}{{\rm d} t} \left( \frac{1}{e^2} \right)
= \frac{a(\cdots)}{e^2} \, ,
\eeq
where the dots indicate that the function `$a$' can depend 
upon parameters of $S_{\rm GR}$.
The path-integral in eq. (\ref{eq:pathint_em}) can be 
re-written by making use of an auxiliary tensor field $B_\mn$.
This is done by introducing a measure for the auxiliary field 
in the path-integral, a determinant factor constructed exclusively from 
metric and modifying the original action by writing 
it in the first order form by making use of auxiliary field. 
Integrating over this auxiliary field, gives back the original path-integral. 
Under this transformation the path-integral becomes,
\bea
\label{eq:dualtrans}
\int {\cal D} A_{\mu} e^{i S_{\rm EM}} = \int {\cal D}B_\mn \, 
{\cal D} A_{\mu} e^{i S_B} \, \left[
\det e^2 G^{\mn\al\bt}
\right]^{1/2} \, ,
\eea
where 
\bea
\label{eq:dualtrans1}
&&
S_B =e^2 \int {\rm d}^4 x \sqrt{-g}  g^{\mu\al} g^{\nu\bt}
B_\mn B_{\al\bt} + \int {\rm d}^4 x \, \ep^{\mn\al\bt}
B_\mn \pt_{\al} A_\bt  \, , \\
&&
\label{eq:dualtrans2}
G^{\mn,\al\bt} = \sqrt{-g} (g^{\mu\al} g^{\nu\bt}- g^{\mu\bt} g^{\nu\al}) \, .
\eea
$G^{\mn,\al\bt}$ is anti-symmetric in $(\mn)$ and 
$(\al\bt)$ and its determinant arises after doing integration over the 
auxiliary field $B_\mn$ and $\bar{\ep}^{\mn\al\bt}$ is a four dimensional tensor density 
of weight $-1$. It should be noted that it is 
flat spacetime partial derivative that enters the second integrand of the 
eq. (\ref{eq:dualtrans1}) (when there is covariant derivative, 
the connection term in the covariant derivative cancels due to the presence of 
$\bar{\ep}^{\mn\al\bt}$, this happens as connection is taken to be
torsionless, situation is different in presence of torsion).

On integrating the $A$ field on the rhs 
of eq. (\ref{eq:dualtrans}) we gets a constraint on the $B_\mn$ field 
$\de \left( \bar{\ep}^{\mn\al\bt} \pt_{\al} B_\mn \right)$. This constraint
appears under the path-integration of the field $B_\mn$. Being a 
$\de$-function, this constraints the $B_\mn$ field to pick up those
configurations which satisfies $ \bar{\ep}^{\mn\al\bt} \pt_{\al} B_\mn=0$.
This is satisfied by solution $B_\mn = \pt_\mu b_\nu-\pt_\nu b_\mu$. The 
field transformation from $B_\mn$ to $b_\rho$ however introduces 
a jacobian constructed purely from partial derivative and Levi-civita 
tensor. This is a trivial normalization constant and wouldn't affect the 
renormalization group study of the couplings. The path-integral 
for the dual-field is given by,
\bea
\label{eq:dualpathint}
Z= C \int {\cal D} g_\mn \, {\cal D} b_{\mu}
e^{i (S_{\rm GR} + \bar{S}_{\rm EM})} 
\left[
\det (e^2 G^{\mn,\al\bt} )\right]^{\frac{1}{2}} 
\, ,
\eea
where $C$ is a constant and $b_\mu$ is the dual 
of $A_\mu$ whose action is given by,
\beq
\label{eq:dualact}
\bar{S}_{\rm EM} = e^2 \int {\rm d}^4 x \, \sqrt{-g} 
g^{\mu\al} g^{\nu\bt} (\pt_\mu b_\nu - \pt_\nu b_\mu) 
(\pt_\al b_\bt - \pt_\bt b_\al) \, .
\eeq
It should be noted that here we have implemented the duality transformation
in the presence of metric. This dual action has also $U(1)$ gauge invariance.
Next we note that the determinant of $e^2 G^{\mn\al\bt}$
which is anti-symmetric in $(\mn)$ and $(\al\bt)$,
is a general coordinate invariant ultra-local $d(d-1)/2 \times d(d-1)/2$ matrix. 
Hence the determinant can only be proportional 
to some power of $\sqrt{-g}$. It is found that
\beq
\label{eq:Gdet_power}
\left( \det \left[e^2 \sqrt{-g} 
(g^{\mu\al} g^{\nu\bt}- g^{\mu\bt} g^{\nu\al}) \right] \right)^{1/2} = 
e^{d(d-1)/2}
\left(
\sqrt{-g} \right)^{(d-1)(d-4)/4} \, .
\eeq
In four dimensions this is a pure number and is equal to $\left(e^2 \right)^3$.
Hence the dual theory path-integral looks exactly the same as the 
original path-integral apart from the overall constant. 
Perturbative renormalization group study of the dual action gives the flow
of the coupling present in the dual field theory. This is given by,
\beq
\label{eq:beta_dual}
\frac{{\rm d}}{{\rm d}t} e^2 = a(\cdots) \, e^2 \, ,
\eeq
where $a$ has exactly the same parameter dependence as in 
Eq. (\ref{eq:e2beta}). 

Compatibility of the two eqs. 
(\ref{eq:e2beta} and \ref{eq:beta_dual}) implies that $a=0$ \cite{NAgauge}. 
This of course comes out at one-loop level explicitly. Here we have not 
introduced any gauge-fixing and corresponding ghost action for the 
analysis. If one uses similar kind of gauge fixing conditions in the 
original and in the dual theory, then it is easy to show that the 
above proof goes through without any modification.
It should be noted that this argument doesn't depend on the
gravity action or the regularization scheme, implying that in
both the $U(1)$ and its dual theory, eq. (\ref{eq:e2beta}
and \ref{eq:beta_dual}) are inferred in any regularization scheme.

In the next section we explicitly show that at one-loop $a=0$
independent of any regularization scheme, gravity action 
and gauge fixing thereof.

\section{Arbitrary Gravity Action}
\label{arbitGR}

Having realized in the previous section that gravitational action play 
no significant role at all in the argument to show $a=0$ to all loops, 
it becomes all the more compulsive to see by checking 
whether it is indeed true in an explicit one-loop computation of the 
gauge beta function for arbitrary gravity action. 
Here in this section we do the same.

This generalization can be achieved in metric theories of gravity by
considering a general action that can be constructed with metric 
and its derivatives. In such general theories of gravity, one can still write
the form of the propagator of the metric fluctuation $h_\mn$ around a flat 
background. This can be done by making use of complete set of spin
projectors for rank-2 tensor field. At this step one can ask
what generality in the gravitational action actually transmits to the metric 
fluctuation propagator around the flat background? It is noticed that
higher-derivative terms of sixth order like $R^3$ and 
$R_\mn{}^{\rho\sg} R_{\rho\sg}{}^{\al\bt} R_{\al\bt}{}^\mn$ don't contribute to 
the propagator but terms like $R\Box R$ (which has same order of 
derivatives) contribute. This observation brings to realization 
that terms of action which are at most quadratic in curvature, contributes
to the propagator of the $h_\mn$ field. Denoting the inverse propagator of the
$h_\mn$ field by $\D_G^{\mn\al\bt}$ as in eq. (\ref{eq:gr_invprop}), the most
general form of it in momentum space can be written as,
\beq
\label{eq:invprop_proj}
\D_G^{\mn\al\bt}(p) = \sum_i \, \tilde{Y}_i(p^2)  P_i^{\mn\al\bt} \, ,
\eeq
where $i=\{2, 1, s, w, sw, ws\}$, $P_i$'s are all the projectors as given 
in appendix \ref{app_proj}. By making use of the properties of these projectors, this
can be inverted to obtain the expression for the most general propagator
around the flat background of the $h_\mn$ field. This can be written as,
\beq
\label{eq:arbitGR_prop}
D^{\mn \rho\sg} = \sum_i \, Y_i(p^2)  P_i^{\mn\al\bt} \, ,
\eeq
where $Y_i$ are propagators for various spin components and are 
related to to $\tilde{Y}_i$'s in the following way,
\bea
\label{eq:invprop_coeff}
&&
Y_2 = \frac{1}{\tilde{Y}_2} \, , \hspace{5mm}
Y_1 = \frac{1}{\tilde{Y}_1} \, , \hspace{5mm}
Y_s = \frac{\tilde{Y}_w}{ \tilde{Y}_s \tilde{Y}_w - \tilde{Y}_{sw} \tilde{Y}_{ws}}\, ,
\notag \\
&&
Y_w = \frac{\tilde{Y}_s}{\tilde{Y}_s \tilde{Y}_w - \tilde{Y}_{sw} \tilde{Y}_{ws}} \, ,
\hspace{5mm}
Y_{sw} = -\frac{\tilde{Y}_{sw}}{\tilde{Y}_s \tilde{Y}_w - \tilde{Y}_{sw} \tilde{Y}_{ws}} \, ,
\hspace{5mm}
Y_{ws} = -\frac{\tilde{Y}_{ws}}{\tilde{Y}_s \tilde{Y}_w - \tilde{Y}_{sw} \tilde{Y}_{ws}} \, .
\eea
In the case of higher-derivative gravity in Landau gauge we have 
only $Y_2$ and $Y_s$. However in a general gauge fixing and for an
arbitrary gravity action all the $Y_i$'s will be present.

Using this as the propagator for the fluctuation field $h_\mn$, one can 
repeat the computation of the charge renormalization at one-loop for 
arbitrary gravity action. This can be done along the same lines as in
section. \ref{ea}. Again we will have two diagrams giving quantum gravity 
contribution to the gauge coupling: tadpole and bubble, except now
we have a general propagator for the $h_\mn$ field. This generality 
also encapsulates within itself the arbitrariness in the choice of 
gauge fixing parameters for the gravity side. 

Following the steps as in subsection. \ref{tad} and \ref{bub}, we compute 
the contributions of the tadpole and bubble diagram respectively. 
Reaching the stage when momentum integration needs to be 
performed (as in eqs. (\ref{eq:tad_mom_exp} and \ref{eq:I0})), 
we argue that as the measure and the integrand is a 
Lorentz covariant quantity, therefore under the momentum integral the tensorial part
of integrand consisting of various combinations of four momenta $p_\mu$
(which is the momentum variable), 
can be replaced by a combination of flat background metric $\eta_\mn$,
with various coefficients, obeying the symmetries of the integrand.
This is then contracted with the vertices to obtain the equations
similar to eqs. (\ref{eq:tadLorInt} and \ref{eq:bub_hdg_momnt}).

In the case of tadpole the vertex $V^{\rho\sg\mn}_{\ta\tau\al\bt}$ 
(given in the appendix \ref{gaugeverprop}) is used to contract with different combination
of $\eta_\mn$ for various spin-projectors to yield,
\beq
\label{eq:tad_cont}
\G_{{\rm Tad}} = -\frac{i}{8e^2} \int\, {\rm d}^dx \bar{F}^a_{\mn}
\bar{F}^{a\mn} \, \int \, \frac{{\rm d}^d p}{\pd}
\sum_i Y_i(p^2) \biggl[
\frac{(d-4)(d-6)}{4} H_i - \frac{d^2 -8d +4}{2} G_i
\biggr] \, ,
\eeq
where the coefficients $H_i$ and $G_i$ corresponding to various 
spin-projectors are given in appendix \ref{momint}. For bubble diagram we use the 
properties of the vertex $V^{a\mn\al\bt}$ which is symmetric is 
first two greek index and anti-symmetric in last two greek index, to obtain the
result. These properties are discussed in appendix \ref{gaugeverprop}. 
The contribution of the bubble diagram is given by,
\bea
\label{eq:bub_cont}
&&
\G_{{\rm Bub}} = \frac{i}{8e^2} 
\int {\rm d}^dx \bar{F}^{a\mn} \bar{F}^a_{\mn}
\int \frac{{\rm d}^dp}{\pd} \sum_i Y_i(p^2)
\notag \\
&&
\times
\biggl[
(d-4)^2A_i + 2(5d-8) B_i -2 (d-4)(d-1) \left( C_i + D_i \right)
+ 4(d^2 -4d +6) E_i \, ,
\biggr]
\eea
where the coefficients $A_i$'s, $B_i$'s, $C_i$'s, $D_i$'s and $E_i$'s
for various spin projectors are obtained in appendix \ref{momint}.
Adding the tadpole and bubble contribution 
after plugging the various coefficients ($A$, $B$, $C$, $D$, $E$, $G$ and $H$) 
we find,
\beq
\label{eq:EA_grav}
\G_{{\rm Grav}} = 
-\frac{i(d-4) \Omega(d) }{8e^2} 
\int {\rm d}^dx \, {\rm tr} F^2 \, ,
\hspace{5mm}
{\rm where \,\,\,}
\Omega(d) = \sum_i \Omega_i
\int \frac{{\rm d}^dp}{\pd} 
Y_i(p^2) \, ,
\eeq
and various coefficients $\Omega_i$'s are just functions of space-time 
dimension $d$ and are given by,
\bea
\label{eq:Ai_d}
&&
\Omega_2 =-\frac{(d-5)(d-2)(d+1)}{4d(d-1)} \, ,
\hspace{5mm}
\Omega_1 =-\frac{d-1}{8d} \, ,
\notag \\
&&
\Omega_s=\frac{d^2 -12d+19}{4d(d-1)} \, ,
\hspace{5mm}
\Omega_{sw}=\Omega_{ws} = \frac{\sqrt{d-1}}{4d} \, .
\eea
From the total contribution of the two diagrams 
obtained in eq. (\ref{eq:EA_grav}) we note that irrespective of the 
propagator of the $h_\mn$ field, this full quantum gravity contribution
to gauge field action is always proportional to $(d-4)$. 
This is independent of the gravity action and the gauge fixing. This
arbitrariness is present in the various $Y_i$'s and doesn't affect the overall
factor of $(d-4)$. It is independent of the regularization scheme as we have not 
performed the momentum integration. In deriving this result 
we only used the Lorentz covariance of the integrand present in the 
loop integral. It should also be noted that this one-loop contribution 
is also independent of the symmetry group of the gauge field. 
Therefore to one-loop we find that $a=0$ for arbitrary gravity action and gauge fixing.
It is also independent of regularization scheme and the symmetry group of the gauge field. 
This further justifies the observation made in sec. \ref{duality} where
it was found that $a=0$ to all loops. Moreover, as the graviton 
propagator is kept arbitrary here, therefore it also implies that
subclass of higher-loop diagrams, which involves graviton self
energy corrections as sub-graphs (corrections 
to graviton propagator in diagrams (c) and (d) of Fig. \ref{fig:1loop}), will also 
cancel each other in four spacetime dimensions. Some of such 
diagrams at two loops are presented in Fig. \ref{fig:AandB}, where 
the graviton propagator has undergone self energy correction.

\section{Discussion and Conclusion}
\label{conc}

In this paper we study gauge fields coupled with higher-derivative gravity,
which has been shown to be perturbatively renormalizable in four 
dimensions \cite{Stelle, Fradkin, Moriya}. However higher-derivative gravity
is plagued with the problem of unitarity. This problem is manifestly apparent 
when the propagator of the metric fluctuation field $h_\mn$ is written 
around a flat background. There it is clearly noticed, the presence of 
negative norm states, namely the propagator of the spin-2 massive 
mode appears with the wrong sign, and thus violates unitarity at tree 
level. This problem of unitarity has recently been studied again in the 
light of quantum corrections \cite{NarainA1, NarainA2}, where it was 
found that in a certain domain of coupling parameter space (which is large 
enough to incorporate the known physics), the one-loop running of gravitational 
parameters makes the mass of spin-2 massive mode run in such a way so that
it is always above the energy scale being probed. In other words under 
quantum corrections the pole in the propagator of the $M$-mode disappears.
This one-loop result was a major boost for further study and led to us investigating 
the coupling of higher-derivative gravity with gauge fields in \cite{NAgauge}.
In \cite{NAgauge} quantum gravity correction to charge renormalization 
was studied, and major results were presented. Here we present the 
details of the computation and discuss the unitarity in the coupled system. 

The path-integral of the coupled system of gauge field with higher-derivative 
gravity is completely well defined in $4-\ep$ dimensions, where it is perturbatively
studied in the dimensional regularization scheme. One-loop computation of the 
charge renormalization shows that there is no quantum gravity correction to the
gauge coupling beta function. This coupled system being renormalizable and 
free from quadratic divergences evades the criticism raised in \cite{Donoghue1}.

Generically, in perturbation theory, the coupled gauge-gravity system will lead
to the gauge coupling beta function consisting of two kinds of terms: `$a$' 
and `$b$' as shown in eq. (\ref{eq:beta_gauge_gr}). It is argued in the 
paper that the `$a$' term is universal to all gauge theories in the sense that it 
is independent of the gauge coupling $e^2$, but depends on the 
gravitational parameters and the information content of the matter sector. 
On the other hand the `$b$' term is not universal. In this paper our intentions were to 
study the nature of this `$a$' term to all loops in four dimensions.

One-loop computation showed no quantum gravity contribution to the gauge 
coupling beta function thereby implying vanishing of `$a$' term in four 
dimensions. Moreover it was found to be zero for all metric theories of 
gravity and furthermore is independent of gauge group, regularization scheme
and gauge fixing condition. Universality of this `$a$' term allowed us to 
isolate it by studying it in the context of simple abelian gauge theories without
matter in four dimensions. However self-duality of abelian gauge theories in 
four dimensions opened another window, which allowed us to look at the 
same problem from a different angle. Motivated by this realization, we performed
duality transformation by making use of an auxiliary field to study the flow 
of the coupling in the dual theory. 
In the case of abelian gauge theories, the dual theory is also abelian (self-duality),
except that the coupling in front of gauge field action changes from $1/e^2$ to $e^2$ respectively.
The beta function of the gauge coupling in the dual theory has the 
same structure as in the case of original theory with the same `$a$'-term,
as the gravity action is same. The only way two equations for running of $e^2$
can be satisfied is when `$a$'-term vanishes. This is an all-loop argument 
and is independent of the gravity action. However 
it is only for higher-derivative gravity that the path-integral of the coupled 
system is renormalizable and well defined. 

We also studied the unitarity of the coupled system of gauge fields
with higher-derivative gravity at one-loop. The beta functions of the 
of the gravitational couplings gets corrected due to the presence of 
gauge fields. It is realized that only positive $\omega$ is allowed
and describes the physical domain. This requirement puts a upper and lower bound
on the renormalization group time $t$ that is allowed by physical domain of $\omega$.
Thus the parameters only flow between
$t_{\rm min}$ and $t_{\rm max}$. For $t<t_{\rm min}$, $\omega$ becomes negative, signaling the 
instability of the vacuum, meaning that one enters a regime
where effects of cosmological constant should be taken into account. 
For $t>t_{\rm max}$ one-loop equation for $\omega$ doesn't have a solution.
This could be an artifact of the one-loop results and might perhaps go away 
when higher loop corrections are incorporated. 

The flow of $G$ is such that it goes to zero in both extremes of RG time $t$.
In $\omega$-space, $G\sim 1/\omega$ in the UV, which is a gauge parameter invariant
result and is also independent of whether gauge fields are present or not.
In the IR however, $G\sim \omega^{7/20}$ and this is gauge parameter dependent, in the
sense that the power with which it goes to zero depends on gauge condition.
The one-loop beta functions of gravitational parameters allow us to compute the 
running of $M^2/\mu^2$, flow of which has a minima at the point $\omega_*$.
Demanding that the value of $M^2/\mu^2$ at minima is greater than one,
assures that throughout the RG evolution it will remain greater than one. 
This can be arranged easily by choosing the initial parameters appropriately.
Once this condition is satisfied, it guarantees that the RG trajectory will be unitary
as in \cite{NarainA1, NarainA2}. Thus the coupled system of gauge field 
with higher-derivative gravity can be made unitary in this domain of coupling
parameter space. Similarly working out the flow of $M^2/\omega \mu^2$ 
using the one-loop beta functions of the gravitational parameters, it is found that it is a
monotonically decreasing function of RG time $t$. Choice of initial parameters 
further decides whether Riccion will be physically realizable or not. Thus
the physical theory will consist of gravitons and gauge bosons either with Riccions or
without them. The flow of $M^2$ and $M^2/\omega$ is such that in low energy
the higher-derivative terms are naturally suppressed favoring the Einstein-Hilbert and 
cosmological constant term. 

Having realized that the coupled system being renormalizable and unitary,
it becomes all the more important to study charge renormalization 
in this system. Finding that $a=0$ to all loops, results in 
dramatic consequences, meaning that photons interacting only with metric fluctuations 
will propagate essentially as free particle at short distances. This is a 
consequence of self-duality property of abelian gauge field action. 
However it should be mentioned that there is indeed finite 
charge renormalization, which decreases the coupling, but this is of
no significance when matter fields are present. In case of 
non-abelian gauge fields, the same phenomenon manifests as there will not 
be any `$a$' term, but the running is controlled by the 
`$b$'-term. 

In the absence of `$a$'-term, the leading contribution to the 
running of gauge coupling comes from the `$b$'-term of the 
beta function. At one-loop the `$b$'-term is completely independent 
of the gravitational parameters {\it i.e.} it is solely a consequence of charge 
interactions alone. However at two-loops things change. In the gauge coupling beta 
function there is not only matter contributions (which do not change the behavior of running
gauge coupling compared to one-loop result), but also contributions which
have both gauge and gravity couplings. These can influence the behavior 
of the running gauge coupling by giving rise to new fixed points etc. This 
can open new avenues to our understanding of gravity 
and gauge field theories. Further higher-loop corrections from gauge fields will also
influence the gravitational sector. It will interesting to know how the 
Landau singularity both in the $U(1)$ gauge theory and 
higher-derivative gravity sector (weyl-square coupling) gets affected 
under the influence of higher-loop corrections? This will be the 
subject of future work.

\bigskip
\centerline{\bf Acknowledgements} 
%
GN would like to thank Romesh Kaul for useful discussions.

\appendix

\section{Projectors}
\label{app_proj}

The metric fluctuation field $h_\mn$ around a general background can be decomposed
into various components by doing a transverse-traceless decomposition. This is equivalent
to doing decomposition of a vector into transverse and longitudinal components.
For the metric fluctuation field $h_\mn$ around a flat background, this decomposition can be
written in momentum space as,
\beq
\label{eq:TTdecomp}
h_{\mn} 
= h^T_{\mn} + \iota \left(q_{\mu} \xi_{\nu} + q_{\nu} \xi_{\mu} \right)
+ \left( \eta_\mn - \frac{q_\mu q_\nu}{q^2} \right) s
+ \frac{q_\mu q_\nu}{q^2} \, w \, ,
\eeq 
where the various components satisfies the following constraints,
\begin{gather}
\label{eq:constraints}
h^T_{\mu}{}^{\mu}=0 \, , \hspace{5mm} 
q^\mu h^T_\mn =0 \, , \hspace{5mm}
q^\mu \xi_\mu=0 \, .
\end{gather}
Here $h^T_\mn$ is a transverse-traceless symmetric tensor, $\xi_\mu$ is a
transverse vector and $s$ and $w$ are two scalars. This decomposition can be neatly written
by making use of flat spacetime projectors, which projects various components of 
$h_\mn$ field into $h^T_\mn$, $\xi_\mu$, $s$ and $w$ respectively. 
These projectors are written in terms of the following two projectors,
\begin{gather}
\label{eq:LTproj}
L_{\mu\nu} = \frac{q_{\mu} \, q_{\nu}}{q^2} \, \ , \hspace{10mm}
T_{\mu\nu} = \eta_{\mn} - \frac{q_{\mu} \, q_{\nu}}{q^2} \, ,
\end{gather}
which are basically the projector for projecting out various components
of a vector field. They satisfy $q^\mu T_\mn=0$ and $q^\mu L_\mn=q_\nu$.
Using them the projectors for the rank-2 tensor field 
can be constructed. These are given by,
\bea
&&
\label{eq:spin2proj}
(P_2)_{\mu\nu}{}^{\alpha \beta}
 = \frac{1}{2} \left[ T_{\mu}{}^{\alpha} T_{\nu}{}^{\beta} + 
T_{\mu}{}^{\beta}T_{\nu}{}^{\alpha} \right] - \frac{1}{d-1} T_{\mu\nu}T^{\alpha\beta} \, ,
 \\
&&
\label{eq:spin1proj}
(P_1)_{\mu\nu}{}^{\alpha \beta} 
= \frac{1}{8} \left[ 
T_{\mu}{}^{\alpha} \, L_{\nu}{}^{\beta}
+T_{\mu}{}^{\beta} \, L_{\nu}{}^{\alpha} 
+ T_{\nu}{}^{\alpha} \, L_{\mu}{}^{\beta}
+T_{\nu}{}^{\beta} \, L_{\mu}{}^{\alpha} 
\right] \, ,
\\
&&
\label{eq:spinsproj}
(P_s)_{\mu\nu}{}^{\alpha \beta} 
= \frac{1}{d-1} T_{\mu\nu} \, T^{\alpha \beta} \, ,
\\
&&
\label{eq:spinwproj}
(P_w)_{\mu\nu}{}^{\alpha \beta} = L_{\mu\nu} \, L^{\alpha \beta} \, .
\eea
The projectors for spin-2, spin-1, spin-s and spin-w form an 
orthogonal set. In the scalar sector there are two more projectors 
(which are not projectors in the strict sense), which along with
spin-s and spin-w projectors form a complete set. They are given by,
\bea
&&
\label{eq:spinswproj}
(P_{sw})_{\mu\nu}{}^{\alpha \beta}  
= \frac{1}{\sqrt{d-1}} T_{\mu\nu} \, L^{\alpha \beta} \, ,
\\
&&
\label{eq:spinwsproj}
(P_{ws})_{\mu\nu}{}^{\alpha \beta}   
= \frac{1}{\sqrt{d-1}} L_{\mu\nu} \, T^{\alpha \beta}  \, .
\eea
The projectors in eqs. (\ref{eq:spin2proj}, \ref{eq:spin1proj}, \ref{eq:spinsproj} and \ref{eq:spinwproj})
forms a complete set in the sense that their sum is unity. 
\beq
\label{eq:projunity}
(P_2)_\mn{}^{\rho\sg} +(P_1)_\mn{}^{\rho\sg}
+ (P_s)_\mn{}^{\rho\sg} + (P_w)_\mn{}^{\rho\sg}
= \de_\mn^{\rho\sg} \, ,
\eeq
where $\de_\mn^{\rho\sg} = 1/2( \de_\mu^{\rho} \de_\nu^\sg 
+ \de^\rho_\nu \de^\sg_\mu)$.
Each of these projectors when act of $h_\mn$ projects out various 
spin components of the tensor field. 
\bea
\label{eq:projcomp}
&&
(P_2)_{\mn}{}^{\rho\sg} \, h_{\rho\sg} = h_{\mn}^T \, ,
\hspace{5mm}
(P_1)_{\mn}{}^{\rho\sg} \, h_{\rho\sg} =  \iota \left(
q_{\mu} \xi_{\nu} + q_{\nu} \xi_{\mu} \right)\, , 
\notag \\
&&
(P_s)_{\mn}{}^{\rho\sg} \, h_{\rho\sg} = (d-1) T_\mn s \, , \hspace{5mm}
(P_s)_{\mn}{}^{\rho\sg} \, h_{\rho\sg} = L_\mn w \, .
\eea
If the projectors $P_2$, $P_1$, $P_s$ and $P_w$ are written as 
$P_{22}$, $P_{11}$, $P_{ss}$ and $P_{ww}$ respectively, then 
all the projectors (including $P_{sw}$ and $P_{ws}$) satisfy the 
following algebra,
\beq
\label{eq:proj_algebra}
P_{ij} P_{mn} = \de_{jm} P_{in} \, ,
\eeq
where $i$, $j$, $m$ and $n$= \{$2$, $1$, $s$, $w$\}.

Now we study the contraction of spin projectors of $h_\mn$ with 
various tensors. For this we first 
consider the trace of projectors $T$ and $L$. This is given by,
\begin{gather}
\label{eq:traceTL}
{\rm Tr} \, L_{\mu\nu} = 1 \, , \hspace{10mm}
{\rm Tr} \, T_{\mu\nu} = d-1 \, .
\end{gather}
Using this and properties of the various projectors satisfying the
algebra in eq. (\ref{eq:proj_algebra}), we obtain the contraction of the 
projectors with various tensors constructed with $\eta_\mn$
and momentum $q_\mu$, which are used in the one-loop computation
of the paper. These tensor are given by,
\bea
\label{eq:vartensors}
&&
\eta_\mn \eta_{\rho\sg}, \hspace{5mm} (\eta_{\mu\rho}\eta_{\nu\sg}
+ \eta_{\mu\sg}\eta_{\nu\rho}), \hspace{5mm}
\frac{q_\rho q_\sg}{q^2} \eta_\mn, \hspace{5mm}
\eta_{\rho\sg} \frac{q_\mu q_\nu}{q^2}, 
\notag \\
&& 
U_{\mn\rho\sg\g\ta} = 
\left(
\eta_{\rho\mu}\eta_{\nu\g}\eta_{\sg\ta} + \eta_{\rho\nu}\eta_{\mu\g}\eta_{\sg\ta}
+\eta_{\sg\mu}\eta_{\nu\g}\eta_{\rho\ta}+ \eta_{\sg\nu}\eta_{\mu\g}\eta_{\rho\ta}
+\eta_{\rho\mu}\eta_{\nu\ta}\eta_{\sg\g} +\eta_{\rho\nu}\eta_{\mu\ta}\eta_{\sg\g}
\right .
\notag \\
&&
\left. +\eta_{\sg\mu}\eta_{\nu\ta}\eta_{\rho\g} + \eta_{\sg\nu}\eta_{\mu\ta}\eta_{\rho\g}
\right)
\eea
And the contraction of these tensors with the various spin-projectors are given 
in Table. \ref{tab3}.

\begin{table}
[h]
\begin{center}
\begin{tabular}{|c|c|c|c|c|c|c|}
\hline
 & $\left(P_2\right)_{\mn\rho\sg}$ & $\left(P_1\right)_{\mn\rho\sg}$ & $\left(P_s\right)_{\mn\rho\sg}$ 
 & $\left(P_{sw}\right)_{\mn\rho\sg}$ & $\left(P_{ws}\right)_{\mn\rho\sg}$ 
 & $\left(P_w\right)_{\mn\rho\sg}$ \\
\hline
$\eta^{\mn}\eta^{\rho\sg}$ & 0 & 0 & d-1 & $\sqrt{d-1}$ & $\sqrt{d-1}$ & 1\\
\hline
$(\eta^{\mu\rho}\eta^{\nu\sg} + \eta^{\nu\rho}\eta^{\mu\sg})$ & $(d+1)(d-2)$
& $\frac{d-1}{2}$ & 2 & 0 & 0 & 2 \\
\hline
$2(q^{\rho}q^{\sg}/q^2) \eta^\mn$
& 0 & 0 & 0 & 0 & $2\sqrt{d-1}$ & 2 \\
\hline
$2 \eta^{\rho\sg} (q^{\mu}q^{\nu}/q^2)$ 
& 0 & 0 & 0 & $2\sqrt{d-1}$ & 0 & 2 \\
\hline
$(q_{\g}q_{\ta}/q^2) U^{\mn\rho\sg\g\ta} $
& 0 & d-1 & 0 & 0 & 0 & 8 \\
\hline
\end{tabular}
\end{center}
\caption{Contraction of projectors with various tensors}
\label{tab3}
\end{table}

\section{Gauge Vertices and propagator}
\label{gaugeverprop}

The one-loop quantum gravity correction to the gauge beta function can be obtained by 
doing second variation of the coupled action with respect to various fields. While the 
gravity part of the action gives the propagator of the fluctuation field $h_\mn$
around the flat background, the gauge field action when varied gives the gauge
field propagator, and various vertices that are used in this one-loop computation.
In the following we will write the various vertices arising from the gauge field action
and the gauge field propagator. After obtaining these vertices we will study the 
contraction of these vertices with various tensor. Such contractions are ultimately
used in the computation of the two diagrams that give the one-loop quantum gravity 
contribution to the gauge beta function.

\subsection{Gauge-Gauge vertices}
\label{gaugegauge}

The second variation of the gauge field action with respect to gauge field,
gives rise to gauge field propagator and gauge-gauge vertices relevant for
the one-loop computation. These vertices arise only for non-abelian gauge 
theories and for abelian gauge theories they are absent. The presence of 
these vertices (which are coming due to self interaction of non-abelian gauge
field) is responsible for asymptotic freedom of gauge coupling in non-gravitational scenarios.
Here we write these vertices for completeness, while it will not be needed 
in our one-loop computation of the quantum gravity correction to 
gauge beta function. 

Expanding the gauge field action eq. (\ref{eq:gaugeact}) up to second order
in the fluctuation field ${\cal A}_\mu$ (where $A_\mu = \bar{A}_\mu + 
{\cal A}_\mu$), we get the expression for the gauge field 
propagator and gauge-gauge vertices. This is written in the 
last line of eq. (\ref{eq:2ndvargauge}). After some manipulations
it acquires the following simplified form,
\bea
\label{eq:simp1}
&&
-\frac{1}{8e^2} \int {\rm d}^dx \biggl[ 
4 \mathcal{A}^{a\mu} \biggl\{
-\eta_{\mu\nu} \delta^{ac} \Box - 2 f^{abc} \eta_{\mu\nu} \left( \partial_{\rho}
\bar{A}^{b\rho} + \bar{A}^{b\rho} \partial_{\rho} \right) 
+ \eta_{\mu\nu} f^{abc} f^{ecb} \bar{A}^d_{\rho} \bar{A}^{e\rho}
\biggr\} \mathcal{A}^{c\nu} 
\notag \\
&&
+ 4 \mathcal{A}^{a\nu} D_{\nu}D_{\mu} \mathcal{A}^{a\mu}
+4 f^{abc} \bar{F}^a_{\mu\nu} \mathcal{A}^{b\mu}\mathcal{A}^{c\nu} 
\biggr]
 \, .
\eea
To this the gauge fixing action given in eq. (\ref{eq:sgf_gauge}) is added.
This gives the gauge field propagator and various vertices relevant 
for one-loop computation.
\beq
\label{eq:simp11}
-\frac{1}{8e^2} \int {\rm d}^dx  \mathcal{A}^{a\mu} \biggl[
\left(\Delta_g \right)^{ac}_{\mu\nu} 
+\left( \Delta_1 \right)^{ac}_{\mu\nu}
+ \left( \Delta_2 \right)^{ac}_{\mu\nu}
+\left(\Delta_3\right)^{ac}_{\mu\nu}
\biggr] \mathcal{A}^{c\nu}
\eeq
where,
\bea
\label{gaugepropver}
&&\left(\Delta_g \right)^{ac}_{\mu\nu}
= 4 \delta^{ac} \left[ -\eta_{\mu\nu} \Box
+ \left(1 - \frac{1}{\xi} \right) \partial_{\mu}\partial_{\nu} \right]
\, , 
\notag \\
&&
\left(\Delta_1 \right)^{ac}_{\mu\nu}
= -8 f^{abc} \left \{ \eta_{\mu\nu} \left( \partial_{\rho}
\bar{A}^{b\rho} + \bar{A}^{b\rho} \partial_{\rho} \right) 
+ \left(1 - \frac{1}{\xi} \right) \left(
\partial_{\mu} \bar{A}^b_{\nu} + \bar{A}^b_{\nu} \partial_{\mu} \right)
\right\}
\notag \\
&& \left(\Delta_2 \right)^{ac}_{\mu\nu}
= 4 f^{abc} f^{ecb} \left\{ \eta_{\mu\nu} \bar{A}^d_{\rho} \bar{A}^{e\rho}
- \left(1- \frac{1}{\xi} \right) \bar{A}^d_{\mu} \bar{A}^e_{\nu} \right\}
\, , 
\notag \\
&&
\left(\Delta_3 \right)^{ac}_{\mu\nu}
= - 8  f^{abc} F^b_{\mu\nu} \, .
\eea
From the above we note that $(\D_g)^{ac}_\mn$ is the inverse 
gauge propagator while $\D_1$, $\D_2$ and $\D_3$ are the gauge-gauge 
vertices. 

\subsection{Gauge-Gravity vertices}
\label{gaugegravity}

At the one-loop level the gauge-gravity vertices arise from the following term 
of the second variation of the gauge field action given in eq. (\ref{eq:2ndvargauge}).
This arises from the variation of gauge field action in eq. (\ref{eq:gaugeact}) 
once with respect to $h_\mn$ and once with respect to ${\cal A}_\rho^a$.
This is given by,
\beq
\label{eq:simp2}
-\frac{1}{8e^2} \int {\rm d}^dx \left( 2 h \eta^{\mu\alpha} \eta^{\nu\beta}
- 8 h^{\mu\alpha}\eta^{\nu\beta} \right)
\left(D_{\mu} \mathcal{A}^a_{\nu} - D_{\nu}\mathcal{A}^a_{\mu}
\right) \bar{F}^a_{\alpha\beta}
\eeq
This can be simplified and re-written in the following form
after doing integration by parts in some terms. This 
vertex in then given by,
\beq
\label{eq:simp3}
= -\frac{1}{8e^2} \int {\rm d}^dx \, 
\biggl[
2 h_{\mu\nu} V^{a\mu\nu\gamma'\gamma} D_{\gamma'} 
\mathcal{A}^a_{\gamma}
- 2 \mathcal{A}^a_{\gamma} V^{a\mu\nu\gamma'\gamma} 
D_{\gamma'}h_{\mu\nu} 
- 2 A^a_{\gamma} D_{\gamma'} V^{a\mu\nu\gamma'\gamma} h_{\mu\nu}
\biggr]
\eeq
where,
\beq
\label{eq:expV}
V^{a\mu\nu\gamma'\gamma} =
\biggl[
\eta^{\mu\nu} F^{\gamma '\gamma} - \left(
\eta^{\mu\gamma '} \bar{F}^{a\nu\gamma}
+ \eta^{\nu\gamma '} \bar{F}^{a\mu\gamma}
- \eta^{\mu\gamma}\bar{F}^{a\nu\gamma '}
-\eta^{\nu\gamma} \bar{F}^{a\mu\gamma '}
\right)
\biggr]
\eeq
From this we note that $V^{a\mn\g,\g}$ is symmetric is $\mn$ and
anti-symmetric in $\g'\g$. In these kind of vertices there is one internal 
gluon line ${\cal A}_\g^a$, one internal metric fluctuation $h_\mn$ line and one
external gluon line, here given by $\bar{F}_\mn$.

\subsection{Gravity-Gravity vertices from gauge action}
\label{gravitygravity}

From the second variation of the gauge field action given in eq. (\ref{eq:2ndvargauge})
we obtain the interaction vertex with two internal metric fluctuation $h_\mn$ line
and two external $\bar{F}$ lines. This is obtained by varying the gauge field 
action given in eq. (\ref{eq:gaugeact}) twice with respect to $h_\mn$ field.
This is given by,
\bea
\label{eq:grgr}
&&
-\frac{1}{8e^2} \int {\rm d}^dx \biggl\{
\left(\frac{1}{4} h^2 - \frac{1}{2} h_{\mu\nu}h^{\mu\nu} \right)
g^{\mu\alpha}g^{\nu\beta}
- 2 h h^{\mu\alpha} g^{\nu\beta} 
+4 h^{\mu\rho}h_{\rho}{}^{\alpha} g^{\nu\beta}
+ 2 h^{\mu\alpha} h^{\nu\beta}
\biggr\} F_{\mu\nu}^a F_{\alpha\beta}^a
\notag \\
&&
=
-\frac{1}{8e^2} \int {\rm d}^dx
\biggl[
h_{\mu\nu} \biggl\{
\frac{1}{4}  \left( \eta^{\mu\nu} \eta^{\rho\sigma}
- \eta^{\mu\rho}\eta^{\nu\sigma} - \eta^{\mu\sigma}
\eta^{\nu\rho} \right) \eta_{\theta\alpha} \eta_{\tau\beta}
- \eta^{\rho\sigma} \eta_{\tau\beta} \left( \delta^{\mu}_{\theta}
\delta^{\nu}_{\alpha} + \delta^{\nu}_{\theta} \delta^{\mu}_{\alpha}
\right) 
\notag \\
&&
+ 2 \delta^{\sigma}_{\alpha} \eta_{\tau \beta}
\left(\delta^{\mu}_{\theta} \eta^{\rho\nu} + 
\delta^{\nu}_{\theta} \eta^{\mu\rho} \right)
+ \frac{1}{2} \left( \delta^{\mu}_{\theta} \delta^{\nu}_{\alpha} 
+ \delta^{\nu}_{\theta} \delta^{\mu}_{\alpha} \right)
\left( \delta^{\rho}_{\tau} \delta^{\sigma}_{\beta}
+ \delta^{\rho}_{\beta}\delta^{\sigma}_{\tau}\right)
\biggr\} h_{\rho\sigma} 
\biggr] F_{\ta\tau}^a F_{\al\bt}^a
\notag \\
&&
= 
-\frac{1}{8e^2} \int {\rm d}^dx \, \, 
h_{\mu\nu} V^{\mu\nu\rho\sigma}_{\theta \tau\alpha\beta}
\bar{F}^{a\theta\tau} \bar{F}^{a\alpha\beta} h_{\rho\sigma} \, ,
\eea
where,
\bea
\label{eq:grvert}
&&
V^{\mu\nu\rho\sigma}_{\theta \tau\alpha\beta}
=\biggl\{
\frac{1}{4}  \left( \eta^{\mu\nu} \eta^{\rho\sigma}
- \eta^{\mu\rho}\eta^{\nu\sigma} - \eta^{\mu\sigma}
\eta^{\nu\rho} \right) \eta_{\theta\alpha} \eta_{\tau\beta}
- \eta^{\rho\sigma} \eta_{\tau\beta} \left( \delta^{\mu}_{\theta}
\delta^{\nu}_{\alpha} + \delta^{\nu}_{\theta} \delta^{\mu}_{\alpha}
\right) 
\notag \\
&&
+ 2 \delta^{\sigma}_{\alpha} \eta_{\tau \beta}
\left(\delta^{\mu}_{\theta} \eta^{\rho\nu} + 
\delta^{\nu}_{\theta} \eta^{\mu\rho} \right)
+ \frac{1}{2} \left( \delta^{\mu}_{\theta} \delta^{\nu}_{\alpha} 
+ \delta^{\nu}_{\theta} \delta^{\mu}_{\alpha} \right)
\left( \delta^{\rho}_{\tau} \delta^{\sigma}_{\beta}
+ \delta^{\rho}_{\beta}\delta^{\sigma}_{\tau}\right)
\biggr\}
\eea
%

\subsection{Properties of Gauge-gravity vertex}
\label{verprop}

Having obtained the various vertices that will be needed for the 
one-loop computation of the quantum gravity contribution to the 
gauge beta function, we set to study their properties by contracting 
them with various tensors. These will be very useful in doing the
computation of the tadpole and bubble diagrams in sections. \ref{tad}
and \ref{bub} respectively. These properties are given as follows,
\bea
\label{eq:vertex_id}
&&
\eta_{\al\bt} V^{c\al\bt\g'\g} = (d-4) \bar{F}^{c\g'\g} \, ,
\\
&&
\eta_{\bt\g'} V^{c\al\bt\g'\g} = -(d-1) \bar{F}^{c\al\g} \, ,
\\
\label{eq:B_tensor}
&&
(\eta_{\rho\al}\eta_{\sg\bt} + \eta_{\rho\bt}\eta_{\sg\al} )\eta_{\g'\ta}
\eta_{\tau\g} V^{c\al\bt\g'\g} V^{c\rho\sg\ta\tau}
= 2 V^{c\al\bt\g'\g}V^{c}{}_{\al\bt\g'\g}
= 2(5d-8)\bar{F}^{a\mn} \bar{F}^a_{\mn} \, ,
\\
\label{eq:C_tensor}
&&
(\eta_{\rho\g'}\eta_{\sg\ta} + \eta_{\rho\ta}\eta_{\sg\g'})
\eta_{\al\bt} \eta_{\tau\g} V^{c\al\bt\g'\g}V^{c\rho\sg\ta\tau}
= -2(d-4)(d-1) \bar{F}^{a\mn} \bar{F}^a_{\mn} \, ,
\\
\label{eq:D_tensor}
&&
(\eta_{\al\g'}\eta_{\bt\ta} + \eta_{\bt\g'}\eta_{\al\ta})
\eta_{\rho\sg} \eta_{\tau\g} V^{c\al\bt\g'\g} V^{c\rho\sg\ta\tau}
= -2(d-4)(d-1) \bar{F}^{a\mn} \bar{F}^a_{\mn} \, ,
\\
\label{eq:E_tensor}
&& U_{\rho\sg\al\bt\g'\ta} \eta_{\tau\g} 
V^{c\al\bt\g'\g} V^{c\rho\sg\ta\tau}
= 4(d^2 -4d +6)\bar{F}^{a\mn} \bar{F}^a_{\mn} \, .
\eea
where $U_{\rho\sg\al\bt\g\ta}$ is defined in eq. (\ref{eq:vartensors}). 
 
\section{Momentum Integrals}
\label{momint}
 
Here we will do the computation of the momentum integrals that are 
witnessed during the process of evaluating the tadpole and bubble
graph in section. \ref{tad} and \ref{bub} respectively. The important 
thing to realize in the evaluation of these integration is that 
the actual momentum integrals are Lorentz covariant quantities. Therefore
any tensor structure appearing in the integrands (made of momentum
integration variable), after the integration, will result in Lorentz covariant 
quantity constructed using background flat spacetime metric $\eta_\mn$.
This is equivalent to replacing the integrand with the same tensor structure constructed 
with flat metric $\eta_\mn$. This simplifies the integration process very much
and in the end leads to evaluation of scalar integrals multiplied by 
various combination of flat metric $\eta_\mn$.
 
In the case of tadpole graph, the following momentum integral is witnessed,
\beq
\label{eq:momint_tad}
\int \, \frac{{\rm d}^dp}{\pd} \, \sum_i \, Y_i(p^2)
\left(P_i\right)_{\mn\rho\sg} (p) \, ,
\eeq
where $i={2,1,s,w,sw,ws}$. Under the $p$-integration we replace $P_i$'s as
\beq
\label{eq:proj_rep_tad}
\left(P_i\right)_{\mn\rho\sg} = H_i \eta_{\mn}\eta_{\rho\sg}
+ G_i(\eta_{\mu\rho}\eta_{\nu\sg} + \eta_{\nu\rho}\eta_{\mu\sg}) \, .
\eeq
Thus the momentum integral after this replacement becomes,
\beq 
\label{eq:TadMom}
\int \, \frac{{\rm d}^dp}{\pd} \, \sum_i \, Y_i(p^2)
\biggl[
H_i \eta_{\mn}\eta_{\rho\sg}
+ G_i(\eta_{\mu\rho}\eta_{\nu\sg} + \eta_{\nu\rho}\eta_{\mu\sg})
\biggr] \, .
\eeq
The reason we choose the above set of combination of $\eta$'s
is because $P_i$'s are symmetric in $(\mn)$ and $(\rho\sg)$.
The only tensors that satisfy these properties are the ones 
given above. Multiplying eq. (\ref{eq:momint_tad} and \ref{eq:TadMom})
with $\eta_{\mn}\eta_{\rho\sg}$ and  $(\eta_{\mu\rho}\eta_{\nu\sg} 
+ \eta_{\nu\rho}\eta_{\mu\sg})$ gives $H_i$'s and $G_i$'s for various spin projectors.
\begin{table}
[h]
\begin{center}
\begin{tabular}[c]{|c|c|c|}
\hline
\rule[-4mm]{0mm}{10mm} \rule[-4mm]{0mm}{10mm}\raisebox{-0.8ex}[0.8ex]{\hphantom{ak}
$P_i$ \hphantom{ak}}& $H_i$ & $G_i$ \\
\hline
\rule[-4mm]{0mm}{10mm} $P_2$ & \hphantom{adk}$-\frac{(d+1)(d-2)}{d(d+2)(d-1)}$\hphantom{adk} 
& \hphantom{akd}$\frac{(d+1)(d-2)}{2(d+2)(d-1)}$\hphantom{akd} \\
\hline
\rule[-4mm]{0mm}{10mm} $P_1$ & \hphantom{adk} $-\frac{1}{2d(d+2)}$ \hphantom{adk}
& \hphantom{akd}$\frac{1}{4(d+2)}$\hphantom{akd} \\
\hline
\rule[-4mm]{0mm}{10mm} $P_s$ & \hphantom{adk} $\frac{d^2-3}{d(d+2)(d-1)}$ \hphantom{adk}
& \hphantom{akd} $\frac{1}{d(d+2)(d-1)}$ \hphantom{akd}\\
\hline
\rule[-4mm]{0mm}{10mm} $P_{sw}$ & \hphantom{adk} $\frac{d+1}{d(d+2)\sqrt{d-1}}$ \hphantom{adk}
& \hphantom{akd}$-\frac{1}{d(d+2)\sqrt{d-1}}$ \hphantom{akd}\\
\hline
\rule[-4mm]{0mm}{10mm} $P_{ws}$ & \hphantom{adk} $\frac{d+1}{d(d+2)\sqrt{d-1}}$ \hphantom{adk}
& \hphantom{akd} $-\frac{1}{d(d+2)\sqrt{d-1}}$ \hphantom{akd} \\
\hline
\rule[-4mm]{0mm}{10mm} $P_w$ & \hphantom{adk} $\frac{1}{d(d+2)}$ \hphantom{adk}
& \hphantom{akd} $\frac{1}{d(d+2)}$ \hphantom{akd} \\
\hline
\end{tabular}
\label{tab2}
\caption{Coefficients for tadpole}
\end{center}
\end{table}

In the same way the momentum integral appearing in the case of 
bubble diagram can be worked out. The momentum integral 
we are interested in evaluating is given in eq. (\ref{eq:I0}). From this
we note that the integrand is 
symmetric in pairs $(\rho\sg)$, $(\al\bt)$ and $(\g'\ta)$, 
and the integral $I(0)$ is Lorentz co-variant. Thus under the momentum 
integration one can replace the tensor $\left(P_i\right)_{\rho\sg\al\bt}
\times p_{\g'}p_{\ta}/p^2$ with the most general tensor constructed 
with $\eta$'s obeying these symmetries. Thus we have,
\bea
\label{eq:Bubproj_transform}
&&
\int \frac{{\rm d}^dp}{\pd} \sum_i Y_i(p^2) 
\left(P_i\right)_{\rho\sg\al\bt} \times \frac{p_{\g'}p_{\ta}}{p^2}
= \int \frac{{\rm d}^dp}{\pd} \sum_i Y_i(p^2) 
\biggl[
A_i \eta_{\rho\sg} \eta_{\al\bt}\eta_{\g'\ta}
\notag\\
&&
+ B_i (\eta_{\rho\al}\eta_{\sg\bt} + \eta_{\rho\bt}\eta_{\sg\al})\eta_{\g'\ta}
+ C_i(\eta_{\rho\g'}\eta_{\sg\ta} + \eta_{\rho\ta}\eta_{\sg\g'})\eta_{\al\bt}
D_i(\eta_{\al\g'}\eta_{\bt\ta} + \eta_{\bt\g'}\eta_{\al\ta})\eta_{\rho\sg}
\notag\\
&&
+ E_i \biggl(
\eta_{\rho\al}\eta_{\bt\g'}\eta_{\sg\ta} + \eta_{\rho\bt}\eta_{\al\g'}\eta_{\sg\ta}
+\eta_{\sg\al}\eta_{\bt\g'}\eta_{\rho\ta}+ \eta_{\sg\bt}\eta_{\al\g'}\eta_{\rho\ta}
+\eta_{\rho\al}\eta_{\bt\ta}\eta_{\sg\g'} +\eta_{\rho\bt}\eta_{\al\ta}\eta_{\sg\g'}
\notag\\
&&
+\eta_{\sg\al}\eta_{\bt\ta}\eta_{\rho\g'} + \eta_{\sg\bt}\eta_{\al\ta}\eta_{\rho\g'}
\biggr)
\biggr] \, ,
\eea
where for each spin-projectors $A$, $B$, $C$, $D$, and $E$ will be
different. They are given in Table. \ref{tab4}.
\begin{table}
[h]
\begin{center}
\begin{tabular}[c]{|c|c|c|c|c|c|}
\hline
\rule[-4mm]{0mm}{10mm} $P_i$ & $A$ & $B$ & $C$ & $D$ & $E$ \\
\hline
\rule[-4mm]{0mm}{10mm}\raisebox{-0.8ex}[0.8ex]{\hphantom{ak} 
$P_2$ \hphantom{ak}} & $-\frac{d+1}{d(d-1)(d+4)}$ & $\frac{(d+1) (d^2+2d-4)}{2 d (d+4)(d+2)(d-1)}$
& $\frac{2(d+1)}{d(d+4)(d+2)(d-1)}$ & $\frac{2(d+1)}{d(d+4)(d+2)(d-1)}$
& $-\frac{d+1}{2(d+4)(d+2)(d-1)}$ \\
\hline
\rule[-4mm]{0mm}{10mm} $P_1$ & $-\frac{1}{2d((d+2)(d+4)}$ & $\frac{1}{4d(d+4)}$
& $-\frac{1}{2d(d+2)(d+4)}$ & $-\frac{1}{2d(d+2)(d+4)}$ 
& $\frac{1}{8(d+2)(d+4)}$\\
\hline
\rule[-4mm]{0mm}{10mm} $P_s$ & $\frac{d^2+4d+1}{d(d+2)(d-1)(d+4)}$
& $\frac{1}{d(d+2)(d-1)(d+4)}$ & $-\frac{d+3}{d(d+2)(d-1)(d+4)}$
& $-\frac{d+3}{d(d+2)(d-1)(d+4)}$ & $\frac{1}{d(d+2)(d-1)(d+4)}$\\
\hline
\rule[-4mm]{0mm}{10mm} $P_{sw}$ & $\frac{d+3}{d(d+2)(d+4)\sqrt{d-1}}$
& $\frac{-1}{d(d+2)(d+4)\sqrt{d-1}}$ & $\frac{-1}{d(d+2)(d+4)\sqrt{d-1}}$ 
& $\frac{d+3}{d(d+2)(d+4)\sqrt{d-1}}$ & $\frac{-1}{d(d+2)(d+4)\sqrt{d-1}}$ \\
\hline
\rule[-4mm]{0mm}{10mm} $P_{ws}$ & $\frac{d+3}{d(d+2)(d+4)\sqrt{d-1}}$
& $\frac{-1}{d(d+2)(d+4)\sqrt{d-1}}$ & $\frac{d+3}{d(d+2)(d+4)\sqrt{d-1}}$ 
& $\frac{-1}{d(d+2)(d+4)\sqrt{d-1}}$ & $\frac{-1}{d(d+2)(d+4)\sqrt{d-1}}$ \\
\hline
\rule[-4mm]{0mm}{10mm} $P_w$ & $\frac{1}{d(d+2)(d+4)}$ & $\frac{1}{d(d+2)(d+4)}$  & $\frac{1}{d(d+2)(d+4)}$ 
& $\frac{1}{d(d+2)(d+4)}$  & $\frac{1}{d(d+2)(d+4)}$ \\
\hline
\end{tabular}
\caption{The coefficients used in bubble diagram}
\label{tab4}
\end{center}
\end{table}



\end{document}